\documentstyle[12pt, epsf]{article}

\begin{document}
\baselineskip=0.7cm
\newcommand{\EQ}{\begin{equation}}
\newcommand{\EN}{\end{equation}}
\newcommand{\EQA}{\begin{eqnarray}}
\newcommand{\EQN}{\end{eqnarray}}
\newcommand{\EQAN}{\begin{eqnarray*}}
\newcommand{\EQNN}{\end{eqnarray*}}
\newcommand{\e}{{\rm e}}
\newcommand{\Sp}{{\rm Sp}}
\renewcommand{\theequation}{\arabic{section}.\arabic{equation}}
\newcommand{\Tr}{{\rm Tr}}
\newcommand{\lpartial}{\buildrel \leftarrow \over \partial}
\newcommand{\rpartial}{\buildrel \rightarrow \over 
\partial}
\renewcommand{\thesection}{\arabic{section}.}
\renewcommand{\thesubsection}{\arabic{section}.\arabic{subsection}}
\makeatletter
\def\section{\@startsection{section}{1}{\z@}{-3.5ex plus -1ex minus 
 -.2ex}{2.3ex plus .2ex}{\large}} 
\def\subsection{\@startsection{subsection}{2}{\z@}{-3.25ex plus -1ex minus 
 -.2ex}{1.5ex plus .2ex}{\normalsize\it}}
\makeatother
\def\thefootnote{\fnsymbol{footnote}}

\begin{flushright}
hep-th/0209251\\
UT-KOMABA/02-09\\
September 2002
\end{flushright}
\vspace{1cm}
\begin{center}
\Large
Holographic Reformulation  
of String Theory\\ on AdS$_5\times$S$^5$ background 
in the PP-wave limit

\vspace{1cm}
\normalsize
Suguru {\sc Dobashi}, 
\footnote{
e-mail address:\ \ {\tt doba@hep1.c.u-tokyo.ac.jp}}
 Hidehiko {\sc Shimada}, 
\footnote{
e-mail address:\ \ {\tt shimada@hep1.c.u-tokyo.ac.jp}}
and 
Tamiaki {\sc  Yoneya}
\footnote{
e-mail address:\ \ {\tt tam@hep1.c.u-tokyo.ac.jp}}
\\
\vspace{0.3cm}

{\it Institute of Physics, University of Tokyo\\
Komaba, Meguro-ku, Tokyo 153-8902}

\vspace{0.6cm}
{\sl To the memory of Prof. Bunji Sakita}

\vspace{0.6cm}
Abstract
\end{center}

The recent proposal on the correspondence between 
the ${\cal N}=4$ super Yang-Mills theory and string theory 
in the Penrose 
limit of the AdS$_5\times$S$^5$ geometry 
involves a few puzzles from the viewpoint of holographic 
principle, especially in connection with the 
interpretation of times. 
To resolve these puzzles, we propose 
to interpret the PP-wave strings 
on the basis of tunneling null geodesics connecting 
boundaries of the AdS geometry. 
Our approach predicts a direct and systematic 
identification of 
the S-matrix of Euclidean string theory in the bulk 
with the short-distance structure of correlation functions 
 of super Yang-Mills theory on the AdS boundary, 
as an extension of the ordinary relation in 
supergravity-CFT correspondence. 
Holography requires an infinite number 
of contact terms for interaction vertices of string field theory and 
constrains their forms in a way consistent with 
supersymmetry.   

\newpage
\section{Introduction}

Recently, the so-called parallel-plane (PP) wave limits 
of AdS geometries have attracted much attention. 
In particular,  the authors of ref. \cite{bmn} 
suggested an intriguing possibility of  
extracting all stringy degrees of freedom in this particular 
limit from the maximally supersymmetric 
Yang-Mills theory in a special large 
$N$ limit, by identifying the 
string oscillation modes 
with local composite operators composed as products  
of large numbers (of order $\sqrt{N}$) of elementary fields. 

In the original AdS/CFT correspondence,  
the basic postulate which realizes the holographic 
principle, namely, 
the correspondence  between bulk gravity (closed string) 
theory and super Yang-Mills theory on the boundary is 
the relation \cite{holography}
\EQ
Z[\phi_0]_{{\rm string/gravity}}=\langle \exp (-\int d^4 x 
\sum_i \phi^i_0(x){\cal O}_i(x) )\rangle_{{\rm ym}}. 
\label{relcorr}
\EN
This identifies the bulk partition function 
with boundary conditions of the form
\EQ
\lim_{z\rightarrow 0} \phi^i(z, x)= z^{4-\Delta_i}
\phi^i_0(x) ,
\label{boundc}
\EN
 for bulk fields $\phi^i$  to be the generating functional 
for correlation functions on the side of Yang-Mills theory 
living on the boundary where $\{\phi^i_0(x)\}$ play the role of the source fields for local operators with definite conformal 
dimensions $\{\Delta_i\}$. 
Although no rigorous derivation  is known, 
the relation (\ref{relcorr}) 
can be interpreted as two different but 
physically equivalent descriptions of 
response of the system consisting 
of a large number ($=N$) of 
D3-branes  
under the influence of probe D3-branes 
which are put outside of the near 
horizon region of the system under 
consideration. \footnote{
For a discussion elucidating this point, we refer the reader to 
\cite{yonereview}. 
}
Namely, the left-hand side is the description from the 
viewpoint of gravity and the right-hand side is the one 
from the effective gauge theory: 
In the former the presence of probe D3-branes is encoded 
as the boundary conditions on the bulk fields, while 
in the latter the influence of the probe 
is represented as the external sources for 
SYM operators which couple with the bulk fields 
at the boundary. 
 
If one trusts this relation,  however, 
the arguments in ref. \cite{bmn} 
which have been followed by many authors raise  
a couple of puzzles. 
First, as elucidated in ref. \cite{gkp} and \cite{frotsey}, 
the string theory on the 
PP wave geometry can be regarded as a semiclassical 
approximation around special null geodesics as the 
trajectories of a string in its massless ground states 
with large angular momenta. 
The null geodesic considered in ref. \cite{bmn} 
traverses a large circle on $S^5$, 
with large orbital angular momentum. 
The affine parameter $\tau$ along such a geodesic can be 
identified with the time parameter
in the global coordinate system of 
AdS$_5$ $(R^4=4\pi g_sN)$, 
\EQ
ds_{ads}^2 =R^2(-\cosh^2 \rho d\tau^2 
+ d\rho^2 + \sinh^2\rho d\Omega_3^2) , 
\EN
which has no horizon. 
In the present paper, we always use 
the space-time coordinates in the 
string-frame metric with the string unit $\ell_s=1$. 
It is easy to check that 
this geodesic never reaches  
the AdS boundary (corresponding to $z\rightarrow 0$ 
using the notation of Poincar\'{e} patch below) and also that it goes, within a finite 
interval with respect to 
global time,  into the horizon of original D3 brane metric, 
described as the Poincar\'{e} patch 
of the AdS space:
\EQ
ds_{P}^2 ={R^2dz^2 \over z^2} +{ dx_3^2-dt^2\over R^2 z^2}.
\EN
Indeed, a typical null geodesic with large angular momentum $J$ of order $(g_{{\rm YM}}^2N)^{1/2}$ 
along an S$^5$ direction whose location in the 
AdS$_5$ space is at $\rho=0$ takes , in terms of the 
Poincar\'{e} coordinate, the form
\EQ
z={1\over \cos \tau}, \quad t=R^2\tan \tau. 
\label{nullgeodesic}
\EN
This means that the PP wave geometry has  
no boundary\footnote{
This does not exclude the possibility of 
 boundaries for the PP-wave geometry 
in different sense. Our point is only that such boundaries, 
however, cannot be easily connected with the 
boundary in the sense of relation (\ref{relcorr}). 
For a (partial) list of other works on holography for
  the PP-wave geometry, see \cite{contrav}.
} at least 
in the original sense of holographic 
correspondence between bulk and boundary as 
signified by eq. (\ref{relcorr}).  
Thus we would not be able to apply the basic holographic 
relation in order to compute correlation 
functions of the Yang-Mills theory using the proposed correspondence 
between string states on the bulk and local operators 
on the Yang-Mills side. 

Second puzzle is that on the Yang-Mills side 
the transverse directions of string include the 
time direction. Of course, the time direction of 
Yang-Mills theory does not  in general coincide with the 
global time of bulk geometry. However, 
when one traverses outside the 
horizon of the D3-brane metric in the large $R$ limit,  
the global time $\tau$ and the time $t$ of D3-branes  
is proportional to each other,  
$R^2\tau \sim t $, along a substantial part of the 
null geodesic 
during a finite time 
(=D3-brane time) interval ($\delta t \ll R^2, |\tau| \ll 1$)
 characterized by the AdS radius 
$R$. 
This is the regime where the trajectory becomes 
closest ($\cos \tau \sim 1$) to the AdS boundary.   
Therefore it is difficult to 
imagine that the Yang-Mills time direction 
turns into one of transverse directions for 
string oscillations, unless one stipulates that the regime 
where a closed string remains outside the horizon 
of D3 geometry could 
somehow be ignored.  In fact, it is even more harder to 
imagine how information in the regime 
where the trajectory remains inside the D3-horizon 
could be related to the dynamics of 
boundary theory. We also emphasize that this problem 
exists  irrespectively of whether we use 
Minkowskian or Euclidean metric (such as 
R$^1\times S^3$) on the boundary. 

In connection with the problem of time, there is another 
puzzle. As was already mentioned above, 
even an infinite 
time interval 
with respect to the D3-brane metric 
corresponds to a finite interval with respect to the affine time by which  the dynamics of strings 
propagating along the 
null geodesics is described. 
Thus the Yang-Mills correlators would then correspond to propagation of strings in the bulk in 
finite intervals with respect to affine time.  
However, in string theory, 
finite time transition amplitudes, in general,  cannot be 
regarded as observables. Only meaningful 
observables are S-matrix elements describing 
transition amplitudes with infinite time intervals. 
It seems difficult to imagine how 
gauge invariant observables on Yang-Mills side 
are related to meaningful observables on 
string-theory side.

It is important to clarify these puzzles: 
Firstly, it is necessary to settle the issues whether and how 
holographic principle can be realized in the 
context of PP-wave geometry. Without establishing 
concrete relation 
between string-theory side and SYM side, we do not 
know how to use the conjectured relation between string theory and gauge theory. 
Secondly, if one wishes to extend the correspondence 
to other interesting cases, for instance, to general D$p$-branes other than 
the D3 case, there is no corresponding thing to 
the global coordinate of the AdS geometry.  Consequently,  
it becomes crucial to have correct interpretations 
of possible holographic relations, if any, 
 by using only the regions 
outside the horizon of D-brane metrics. 
For instance, such possibility for 1+0 dimensional 
case may provide  means of identifying the 
stringy degrees of freedom in Matrix theory.  

In the present paper, we propose a new approach 
in which 
the basic  idea behind the holographic relation (\ref{relcorr}) 
is kept as the fundamental  premise for all our arguments.  
We show that this is indeed possible and that it 
predicts  a direct relation between string S-matrix 
in Euclidean formulation and 
gauge invariant Yang-Mills correlators, 
as an extension of the relation (\ref{relcorr}) in a 
suitable short-distance limit. 
This also provides a natural explanation from holography 
for the conjecture made in ref. \cite{constetal},  
and 
shows how the latter conjecture should be 
generalized to higher-point interactions. 

The plan of our paper is as follows. 
In the next section, we argue that 
semi-classical particle picture for the 
basic relation (\ref{relcorr})  should be based on tunneling null geodesics, instead of the 
real geodesics such as (\ref{nullgeodesic}). 
It is shown that by doing this all of the above puzzles 
related to times are 
naturally resolved. 
In section 3, we demonstrate that the large $R$ limit of 
the world-sheet dynamics about the tunneling 
null geodesics can be described by 
the similar (but not identical) 
free massive two-dimensional field theory  
as the ordinary one based on a real null geodesic. 
This leads to a natural identification 
between the Euclideanized string S-matrix and 
the correlators of the operators identified 
in \cite{bmn}. We believe that our proposal 
essentially solves the issue of holography 
for the PP-wave geometry. This is discussed in 
 section 4. We show that the identification 
of the string S-matrix and the correlators 
provides a natural explanation of the 
conjectured relation between 
3-point vertex of string field theory and 
the coefficients of operator-product expansion (OPE). 
Furthermore, our ansatz provides 
a basis for generalizing 
the correspondence to higher-point string 
amplitudes. 
We show, at the level of tree approximation, 
that holography essentially 
fixes the higher-point vertices of the type $1\rightarrow n$ or 
$n\rightarrow 1$  
of string-field theory in terms 
of OPE coefficients. Furthermore, it turns out that 
the constraints required by holography 
quite nicely conform to supersymmetry. 
 The concluding section 5 will be devoted to 
further remarks. In Appendix, we briefly 
discuss the coordinate transformation 
associated with our Euclidean PP-wave limit, 
providing an independent derivation of a part 
of the results in section 3. 

\vspace{0.4cm}
\section{Bulk-boundary correspondence and 
tunneling null geodesics}
\setcounter{equation}{0}

In order to motivate our arguments 
in later sections, let us 
start from considering briefly the propagation of a 
massless scalar field in the background 
of AdS$_5\times $S$^5$ geometry. 
For definiteness, we first assume Minkowskian metric. 
Then the field equation takes the form 
\EQ
\Big(z^2 \partial_z^2 -3 z\partial_z + R^4z^2 \omega^2
-J(J +4)\Big)\phi(z)=0 .
\label{swaveeq}
\EN
We have already factorized the 
angular dependence along the large circle 
(parametrized by angle $\psi$ along a 
great circle of S$^5$) and  the ${\bf R}^{3,1}$ part by assigning, respectively,  a definite 
large angular momentum $-i\partial_{\psi} 
\rightarrow J$ and 
a definite time-like momentum $\partial_4^2 
\rightarrow \omega^2 $ ($> 0$). 
 Now, in the approximation of local-field theory, what corresponds to the string 
picture is the point-particle approximation to the wave equation (\ref{swaveeq}).  In particular, the 
PP-wave limit corresponds to 
$J \propto R^2 \gg 1$.  It is then natural to 
treat the wave equation
 by the WKB approximation 
along the trajectory of a particle. In the above
 factorized form, the WKB approximation is 
described as a simple one-dimensional problem 
 by avoiding technical 
complications associated with higher-dimensional 
configuration space. 

In discussing the 
correlation functions, we are interested in the bulk-boundary 
propagator, satisfying the boundary condition 
(\ref{boundc}) when the bulk-point  approaches to 
the boundary, $z\rightarrow 
0$.  Let us examine how this form emerges 
in the WKB approximation. 
Using the ansatz (${\cal N}=$normalization factor)
\EQ
\phi(z) \sim {\cal N}A(z)\exp iS(z) ,
\EN
where $S$ and $A$ are assumed to be of order 
$J$ and of order one, respectively, we have 
\EQ
z^2\Big({dS\over dz}\Big)^2-R^4z^2\omega ^2 
+J^2=0, 
\label{wkb1}
\EN
\EQ
A(z) =J^{1/2}z^{3/2}\Big({dS\over dz}\Big)^{-1/2}
\exp\Big[-2iJ \int {dz\over z^2}\Big({dS\over dz}\Big)^{-1} 
\Big].
\label{wkb2}
\EN
    From (\ref{wkb1}) which is nothing but the null condition, 
it is evident that there is no particle trajectory approaching to the 
boundary $z=0$, as long as we assume real $S$ since 
it requires $z^2\ge J^2/(\omega^2R^4)$. 
This is of course the origin of one of the puzzles we have 
discussed above. If we still wish to use particle picture, 
we are led to consider 
tunneling wave functions by assuming 
purely imaginary action, $S\rightarrow iS_E, \, \, \phi(r) \rightarrow 
NA(z)\exp -S_E(z)$. 
Then, (\ref{wkb1}) is replaced by 
\EQ
z^2\Big({dS_E\over dz}\Big)^2= 
J^2(1-{z^2\omega ^2R^4 \over J^2}) ,
\EN
and we can now take the near boundary limit $z\rightarrow 0$ limit, 
\EQ
S_E(z) \sim \pm J \log z, \quad A(z) \sim z^{2\mp 2} ,
\label{wkbsol}
\EN
which reproduces the behavior (\ref{boundc}), provided 
\EQ
\Delta =J+4 \quad \mbox{or} \quad  -J ,
\EN
corresponding to the well known mass-dimension relation 
$m^2=J(J+4)=\Delta(\Delta -4)$ for scalar field. 
The former positive solution $\Delta =J+4$  
represents the {\it non}normalizable wave function 
which is responsible for the relation 
(\ref{relcorr}) and (\ref{boundc}). 

The negative solution would correspond to the 
tail of the normalizable wave function, 
which should perhaps be related to 
the ordinary real null geodesics described by 
(\ref{nullgeodesic}). This might be a possible 
hint on how real geodesics may be related in some 
indirect way to the physics occurring at the 
boundary. However, 
in the present paper we do not pursue this possibility. 

This simple exercise clearly shows that in order to 
reconcile the PP-wave limit with the holographic 
relation (\ref{relcorr}), we should consider 
tunneling trajectories by assuming purely 
imaginary momentum along the $z$ direction. 
In fact, because of the existence of the potential barrier 
associated with large angular momentum, it was 
 obvious from the beginning that the boundary-bulk connection for large 
KK momentum along S$^5$ must be 
actually a tunneling phenomenon.   
Let us therefore determine the trajectory of an  
imaginary null geodesic under this assumption. 
By replacing $dS_E/dz$ by the $z$-momentum 
$P_z= Jz^{-2}dz/d\tau$
with $\tau$ being the 
affine parameter, 
we solve 
\EQ
{dz\over d\tau}=\pm z\sqrt{1-{R^4\omega^2 z^2\over J^2}} ,
\EN
and obtain
\EQ
z={J \over R^2\omega \cosh \tau} .
\label{tunnelgeodesic}
\EN
(Our convention for other momenta is  $
J=P_{\psi}=Jd\psi/d\tau, \omega=-P_t=
J(R^4z^2)^{-1}dt/d\tau$. See also section 3.)
This describes a tunneling path, `tunneling null 
geodesic',  which starts from the boundary 
$z=0$ at $\tau=-\infty$, goes into the AdS space 
till the turning point $z=J/R^2\omega$,  and finally 
comes back again 
to the boundary at $\tau =+\infty$. 
Also using the momentum 
$P_t=-J(R^4z^2)^{-1}dt/d\tau=\omega$ along the time 
(target time) 
direction, we obtain 
\EQ
r={J\over \omega}\tanh \tau. 
\EN
 Here, 
in fact, since we made use of purely imaginary 
affine parameter $\tau \rightarrow -i\tau$ but 
simultaneously kept the 
energy $\omega$ and angular momentum $J$ 
fixed, the target time and angle variables 
must also be understood to be 
Wick-rotated to the imaginary axes $t\rightarrow -ir, 
\psi \rightarrow -i \psi$.  
The distance $|r|$ with respect to Euclidean target time $r$ 
 on the 
boundary  is given by 
\EQ
|r|=2{J\over \omega}. 
\label{ymdistance}
\EN
Note that with the above double Wick rotation in the 
target space, 
the ordinary Minkowski null geodesic and 
the tunneling one are connected by the exchange 
$\tau \leftrightarrow -i\tau$, as  is easily seen 
by comparing the real and tunneling solutions. 

These behaviors represent the classical particle 
picture for the boundary-to-boundary propagator 
which is relevant for computing 2-point correlation functions 
on the basis of the relation (\ref{relcorr}). 
We can indeed determine the factorized wave function 
$\psi(x)=f(t)=\int d\omega 
\tilde{f}(\omega)\e^{-i\omega t}$ corresponding to the propagator to the leading order with respect to large $J$. 
Assuming large $t$, the distance 
with respect to $r$ is determined by 
the saddle-point equation
\EQ
{\partial \over \partial \omega}(-\omega r 
+ \log \tilde{f}(\omega))=0, 
\EN
after the Wick rotation.  Comparing with (\ref{ymdistance}), 
we get 
\EQ
\tilde{f}(\omega)=\omega^{2J} ,
\label{omegaamp}
\EN
up to a normalization constant which is independent of $\omega$.  This gives the correct result for two-point correlators ($\propto r^{-2\Delta}$) of operators with conformal 
dimension $\Delta \sim J$ to the leading order 
in the  large $J$ limit.

It is now manifest that in our picture the natural 
time direction along the geodesic is  asymptotically orthogonal  to the base space (${\bf R^{3,1}}\rightarrow 
{\bf R}^4$) of the boundary theory, 
\EQ
z \rightarrow {2J\over R^2\omega}\e^{-|\tau|} 
\quad 
\mbox{as}  \quad \tau \rightarrow \pm \infty. 
\label{ztaurelation}
\EN
This also exhibits that the affine time $\tau$ 
can be identified with the renormalization 
scale parameter of boundary theory, in conformity 
with the identification of its conjugate 
energy $\partial_{\tau} \propto \pm
\Delta $ ($\tau \rightarrow \pm \infty$) with the conformal dimension 
of the boundary operators, apart from the 
angular part of the wave functions which 
contribute $J$ for $\partial_\tau$:
$\e^{iJ\psi}\rightarrow 
\e^{J\tau}$.   If we 
 identify the UV cutoff parameter as $z\sim 1/\Lambda$ 
for $z\rightarrow 0$, the WKB solution (\ref{wkbsol}) indicates that 
 the asymptotic form of the 
boundary-to-boundary amplitude 
is proportional to 
\EQ
\Lambda^{2J} \propto \omega^{2J} \e^{2JT}=\tilde{f}(\omega)
\e^{2JT} ,
\label{clamp}
\EN
in conformity with (\ref{omegaamp}), 
where $2T$ $(\rightarrow \infty)$ is the time interval, 
with respect to the affine time $\tau$ along the 
null geodesics,  of propagation from 
boundary-to-boundary.   Thus the two-point correlators  
can essentially be identified with the classical part of 
two-point 
S-matrix elements computed  
along the tunneling null geodesics. If we combine the 
radial and angular parts  together, the $J$ 
dependence cancels as $ \e^{2\Delta T-2JT} $ to 
the leading order of the present approximation.  

It should also be 
noted that since both $J$ and $\omega$ are 
conserved, they can be assumed to be 
always proportional to each other 
$J/\omega =$const. This leads to the fact that classical 
tunneling paths with various different 
angular momenta can all be identified with one 
and the same trajectory. In other words, 
we can assume the single tunneling trajectory 
for interacting particles around this background:
Particles  can split and join but move along the 
same trajectory at least in the classical approximation. 

Now it is more or less clear that
 all the puzzles we have discussed 
may be solved if we decide to formulate 
string theory in the PP-wave limit 
on the basis of the above tunneling 
null geodesic. 
Our next task is then to examine whether  string theory 
around the tunneling null 
geodesic can be formulated in a well defined way. 

\vspace{0.4cm}
\section{String theory around the tunneling null geodesics}
\setcounter{equation}{0}

\subsection{Derivation of the string action}
We  now study the type IIB string theory 
around the tunneling null geodesic (\ref{tunnelgeodesic}). 
 The bosonic world-sheet action is in the standard conformal 
gauge for the world-sheet metric 
($\ell_s=1$) 
\EQ
S_b = {R^2\over 2\pi}\int d\tau \int_0^{2\pi \alpha} d\sigma 
\, 
{1\over 2}\Big[
z^{-2}(\partial z)^2 + 
z^{-2}(\partial x_4)^2 
-\cos^2 \theta (\partial \psi)^2
+(\partial \theta)^2 +
\sin^2\theta (\partial \Omega_3)^2
\Big], 
\EN
where we have performed the double Wick rotation 
for the target time and angle $\psi$ as discussed 
in the previous section. Hence, four-dimensional 
kinetic term  
$(\partial x_4)^2$ has Euclidean signature, 
and the kinetic term corresponding to the 
angle $\psi$ has negative sign. In the original signature, 
the $S^5$ metric is written as 
\[
ds_5^2=R^2 (  \cos^2 \theta d\psi^2
+d\theta^2 +
\sin^2\theta d\Omega_3^2) .
\]
Also we have rescaled the ${\bf R}^4$ coordinates 
$x_4^i \rightarrow R^2 x_4^i$  
in order to eliminate  the factor 
$R^4$ such that the metric has $R^2$ as 
the global normalization factor. 
The fermionic part will be considered later. 

In the large $J \propto R^2$ limit, it is natural to choose the 
string-length parameter to be 
\EQ
\alpha = J/R^2 \, \, \, (>0) ,
\EN
since it can be treated as a continuous parameter 
which is strictly conserved. Furthermore, it is convenient to 
set, in the notation of the previous section, 
\[
J=R^2 \omega ,
\] 
 or equivalently $\alpha = \omega$. 
Then the classical background 
we consider is 
\EQ
z={1\over \cosh \tau}\equiv z_0, \quad r =\tanh \tau
\equiv r_0 ,\quad \psi =\tau=\psi_0, 
\label{clsol}
\EN 
with all other coordinates being zero, 
where we have denoted the direction of the trajectory 
projected on the boundary by $r$ (Euclidean 
target time), following the 
convention of the previous section. This is of course the 
classical solution to the above action. 

Since the world-sheet time is also Wick-rotated, the 
signature of the world-sheet parametrization is 
Euclidean. In ordinary field theory, the existence 
of the negative metric would then make  the theory 
ill-defined. In string theory, this difficulty is saved by the 
Virasoro constraints which can serve to eliminate 
the negative metric term. Apart from the fermion 
contribution, we have 
\EQ
z^{-2}
(\dot{z}^2-z'^2+ \dot{x}_4^2-x_4'^2)
-\cos^2\theta\,  (\dot{\psi}^2 - \psi'^2)+
\dot{\theta}^2 -\theta'^2 
+\sin^2 \theta\,  (\dot{\Omega_3}^2 
-\Omega_3'^2)=0 ,
\EN
\EQ
z^{-2}(\dot{z}z'+ \dot{x}_4 x_4') 
- \cos^2 \theta \, \dot{\psi}\psi' 
+\dot{\theta}\theta' + \sin^2 \theta \, \dot{\Omega}_3
\Omega_3'
=0 .
\EN
For the classical solution (\ref{clsol}), the first 
is nothing but the null condition and the second 
is trivially satisfied. 

We now expand the action and the Virasoro 
constraints around the 
classical solution 
by making the usual decomposition 
\EQ
z=z_0+{1\over R}z^{(1)} + {1\over R^2} z^{(2)} + 
\cdots, \quad 
r=r_0 +{1\over R}r^{(1)} + {1\over R^2} r^{(2)} +
\cdots ,
\EN
and similarly for other components whose classical parts  
are zero.  
The second order action is then 
\[
S_b^{(2)}={1\over 4\pi}\int d\tau \int_0^{2\pi \alpha}
d\sigma \Big[
{1\over z_0^2}(\partial z^{(1)})^2 
+{1\over z_0^2}(\partial x_4^{(1)})^2
+ {3(z^{(1)})^2\over z_0^4}(\dot{z_0}^2 
+ \dot{r}_0^2 )
\]
\EQ
-4 {z^{(1)}\over z_0^3}(\dot{z}_0\dot{z}^{(1)}
+ \dot{r}_0\dot{r}^{(1)})
-(\partial \psi^{(1)})^2 
+ \dot{\psi}_0^2 (y_4^{(1)})^2
+(\partial y_4^{(1)})^2 
\Big].
\label{secondordaction}
\EN
 Here we denoted the four of the first order coordinates 
of S$^5$ other than the $\psi^{(1)}$ collectively by $y_4^{(1), i}$ $(i=1, \ldots, 4)$. In the large $R$ limit,  3rd and higher order terms does not contribute. 
The first order Virasoro constraints 
are 
\EQ
-{1\over z_0^3}z^{(1)} (\dot{z}_0^2 + \dot{r}_0^2)
+{1\over z_0^2}z^{(1)} (\dot{z}_0 \dot{z}^{(1)} + \dot{r}_0
\dot{r}^{(1)})
-\dot{\psi}_0\dot{\psi}^{(1)}=0 ,
\label{vira1}
\EN
\EQ
{1\over z_0^2}(\dot{z}_0(z^{(1)})'
+ \dot{r}_0 (r_0^{(1)})' )
-\dot{\psi}_0 (\psi^{(1)})'=0 ,
\label{vira2}
\EN
which can serve to eliminate $\dot\psi^{(1)}, 
(\psi^{(1)})'$.  Fermionic degrees of freedom 
does not contribute to this order, since the classical 
solution for fermions is simply zero.  
It is easy to see that the second order Virasoro constraints 
can similarly determine the second order fluctuations 
$\dot{\psi}^{(2)}, (\psi^{(2)})'$ in terms of other 
variables just as the longitudinal excitations 
are eliminated by the Virasoro constraints 
in flat space-time. Since they do not appear in the 
second order action (\ref{secondordaction}), 
it is sufficient for our purpose 
to consider only the first order constraints.    

Using (\ref{vira1}) and (\ref{vira2}) and the explicit 
form of the classical solution, 
the second order action is rewritten as  
\[
S^{(2)}_b =
{1\over 4\pi}\int d\tau \int_0^{2\pi \alpha}
d\sigma
\Big[
(\partial z)^2+z^2+\sinh^2\tau (\partial r)^2 +
\cosh^2 \tau (\partial x_3)^2 
\]
\EQ
+2\cosh \tau r \dot{z} + 2\sinh\tau  zr +
2\sinh \tau \partial z \partial r 
+ (\partial y_4)^2 + y_4^2 
\Big] ,
\EN
where three coordinates of {\bf R}$^4$ other than `time' 
$r$ are separated and collectively denoted by 
$x_3^i$ $(i=1, 2, 3)$.  Note also that we have dropped 
the superscript $(1)$ for notational brevity :
$z^{(1)} \rightarrow z$, {\it etc} . 
The explicit $\tau$ dependence of this action 
can be eliminated, together with cross terms, 
 by making the following 
field redefinition 
\EQ
r + {z\over \sinh \tau}
 \rightarrow {r \over \sinh \tau}, \quad x_3^i \rightarrow 
{x_3^i\over \cosh \tau}. 
\EN
By this redefinition, the variable $z$ is also eliminated. 
By renaming $(x_3^i, r)$ as $x_4^i$ $(i=1, \ldots, 4)$, 
the bosonic action then takes the simple form
\EQ
S^{(2)}_b =
{1\over 4\pi}\int d\tau \int_0^{2\pi \alpha}
d\sigma
\Big[
(\partial x_4)^2 + x_4^2 + (\partial y_4)^2 + y_4^2
\Big]. 
\label{finalbosonicaction}
\EN
Thus we arrived at the free massive world-sheet 
theory with Euclidean metric. 
It is also possible to arrive at the same result 
by directly making suitable coordinate 
transformation. See Appendix. 

Let us now turn to the fermion action. In our quadratic 
approximation which becomes exact in the limit $R^2\rightarrow \infty$, the GS action using Euclidean signature (because of our Wick-rotation 
in $\tau$) for world-sheet metric
is 
\EQ
S^{(2)}_f 
= {i\over 2\pi}\int d\tau \int_0^{2\pi \alpha}
d\sigma \, 
(\delta^{IJ} \delta^{ab}
-i\epsilon^{ab}s^{IJ})\theta^I\Gamma_0\rho_a D_b \theta^J
,
\EN
with $I, J =1, 2$ and 
$
s^{IJ} =\pmatrix{ 1 & 0 \cr 
                                   0 & -1 \cr}. 
$
Here,  $\rho_a$ is the projection of the Gamma matrices 
along the tangent space of the world sheet
\EQ
\rho_a=\Gamma_A E^A_M\partial_a x^M, 
\EN
and $D_a$ is the similar projection of  the extended 
covariant derivative 
$D_M^{IJ}=\delta^{IJ}{\cal D}_M 
 -\epsilon^{IJ}{1\over 8\cdot 5!}F_5
\Gamma^5\Gamma_M $
including the contribution from the self-dual 
5-form RR field strength $F_5=\epsilon_5 + \ast \epsilon_5$, 
\[
D^{IJ}_a = \partial_a x^M D^{IJ}_M ,
\]
which leads to 
\EQ
D_a^{IJ} = {\cal D}_a \delta^{IJ}+{1\over 2}
\epsilon^{IJ} \Gamma_{\ast}
\rho_a , \quad \Gamma_{\ast} =i\Gamma_{0123Z}, \quad 
(\Gamma_{\ast}^2=1)
\EN  
\[
{\cal D}_a=\partial_a + 
{1\over 4}\omega_M^{AB}\partial_ax^M\Gamma_{AB},
\]
where ${\cal D}_a=\partial_ax^M {\cal D}_M$ denotes the standard 
spinor covariant 
derivative along the world-sheet. 
The target space-time geometry is  the 
AdS$_5\times$S$^5$ with the 
double Wick rotation being assumed. For the corresponding 
discussion for the real null geodesic, see \cite{mettsey} and 
references therein.  Hence, 
the Lorentz indices of the Dirac Gamma 
matrices must be understood with 
additional pure-imaginary factor 
when $A, B, \ldots =0$ and $A, B, \ldots =\psi$. 
 It should be noted that the self-duality condition 
for 5-form field strength is preserved under the 
double Wick rotation and that, comparing with the 
Minkowski (with respect to 
both world-sheet and target space-time) action,  
the Levi-Civita symbols $\epsilon^{ab}, \epsilon^{IJ}$ 
acquire pure-imaginary factor $i$. 
The pure-imaginary factor for $\epsilon^{IJ}$ 
originates  from the RR-field strength 
$F_5 \rightarrow iF_5$. 
The spinor coordinate $\theta^I$ $(I=1,2)$ 
are Majorana-Weyl spinor in the Minkowski metric. 
After the double Wick rotation, their components 
must in general be regarded as complex with the {\it same}  
number of independent components {\it as} in the case 
of ordinary Minkowski signature.  Note also that 
the action does not involve complex conjugation 
and hence that this does not cause 
any trouble in Wick-rotating the fermion coordinates. 
In general 
path-integral formalism, as is well known,  it is not 
meaningful to consider hermitian conjugation for  
fermions in Euclidean case in the same sense as in  the 
ordinary Minkowski case. 

Using the space-time metric along the classical solution 
(\ref{clsol}), we find that the fermionic action is written as 
\EQ
S_f^{(2)}=
{i\over 2\pi}\int d\tau \int_0^{2\pi \alpha}
d\sigma \, 
\Big[
\theta^I \Gamma_0\rho_{\tau}(\partial_{\tau}
+iz{1\over 2}\Gamma_{Z0})\theta^I
-is^{IJ}\theta^I\Gamma_0\rho_{\tau}\partial_{\sigma}
\theta^J
+{1\over 2}\epsilon^{IJ}\theta^I\Gamma_0
\rho_{\tau}\Gamma_{\ast}\rho_{\tau}
\theta^J 
\Big]
,
\label{fermionaction}
\EN  
where 
\EQ
\rho_{\tau}=
-z\sinh \tau \Gamma_Z +iz\Gamma_0 + i\Gamma_{\psi}
,
\EN
which satisfies  
\EQ
\rho_{\tau}^2=0 , \quad [{\cal D}_{\tau}, \rho_{\tau}]=0 , 
\quad ({\cal D}_{\tau}=\partial_{\tau}
+iz{1\over 2}\Gamma_{Z0}).
\EN
The $\kappa$-symmetry transformation  in the present approximation 
can be expressed as 
\EQ
\delta_{\kappa}\theta^I = \rho_{\tau}\kappa^I ,
\EN
under which (\ref{fermionaction}) is manifestly invariant.  
To simplify the fermion action further, let us make 
the following field redefinition
\EQ
\theta^I \rightarrow \e^{-i{\beta\over2}\Gamma_{Z0}}
\theta^I , \quad \theta^I \Gamma_0
\rightarrow \theta^I \Gamma_0 \e^{+i{\beta\over2}\Gamma_{Z0}} ,
\EN
such that 
\EQ
\e^{i{\beta\over2}\Gamma_{Z0}}
{\cal D}_{\tau} \e^{-i{\beta\over2}\Gamma_{Z0}}
=\partial_{\tau}, \quad 
\e^{i{\beta\over2}\Gamma_{Z0}}\rho_{\tau}
\e^{-i{\beta\over2}\Gamma_{Z0}}=\Gamma_Z +i\Gamma_{\psi}\equiv \sqrt{2}\Gamma_-. 
\EN
The left hand side of these equations are 
chosen in such a way that they coincide with their 
respective asymptotic forms in the 
limit $\tau \rightarrow -\infty$. This is achieved by setting 
\EQ
\sin \beta = {1\over \cosh \tau}. 
\EN
To fix the $\kappa$-symmetry gauge, we adopt the gauge 
condition, 
\EQ
\Gamma_+ \theta^I =0,  \quad \Gamma_+\equiv
(\Gamma_Z -i \Gamma_{\psi})/\sqrt{2}=\Gamma_-^{\dagger} . 
\EN
We remind the reader again 
that the above field redefinition and 
gauge condition are allowed because the 
fermionic coordinates are treated  as complex with 
the same number of degrees of freedom as in the 
ordinary Minkowski case and also that the action does not 
involve complex conjugation.

Putting all these together, our final form of the second-order fermionic 
action is 
\EQ
S_f^{(2)}=
{i\over 2\pi}\int d\tau \int_0^{2\pi \alpha}
d\sigma \, 
\Big[
\theta^I\Gamma_0\Gamma_-\partial_{\tau}
\theta^I
-is^{IJ}\theta^I\Gamma_0\Gamma_- \partial_{\sigma}
\theta^J
-i\epsilon^{IJ}\theta^I\Gamma_0
\Gamma_-\Pi
\theta^J\Big]
,
\label{fermionaction}
\EN
with 
\EQ
\Pi = i\Gamma_{0123}, \quad \Pi^2 =1 , \quad 
[\Pi, \Gamma_{\pm}]=0=\{\Pi, \Gamma_0\}, \quad \Pi^T=-\Pi, \quad \Pi^{\dagger}=\Pi.
\EN
This is the expected form of Euclidean action 
for free massive fermions as the
 super-partner of the bosonic action 
(\ref{finalbosonicaction}). 
 Because of  the coupling of RR fields in the mass term, 
the manifest global bosonic 
symmetry of the total action 
$S=S_b^{(2)} + S_f^{(2)} $ is naively 
SO(4)$\times$SO(4)$\times Z_2$, 
$Z_2$ being associated with the discrete symmetry 
interchanging two bosonic SO(4) directions. 
 Unlike the usual treatment based on the 
real null geodesics \cite{met}, four 
$(i\times 0, 1, 2, 3)$ of the eight transverse directions 
 can be identified manifestly with the base space of the 
boundary conformal theory of Euclidean signature including 
time direction.\footnote{
In our convention, the other 4 transverse directions are 
$(5, 6, 7, 8)$. The remaining longitudinal (`light-like') directions are 
$z=4, \psi=i\times 9$. } 
For this reason,  our action is {\it not} exactly the 
Wick-rotated version of the action 
which we would obtain in the PP-wave limit from
 the real null geodesics, although the formal structure 
of both actions is very similar to each other. For example, 
our $\Gamma_{\pm}=
(\Gamma_Z\mp i\Gamma_{\psi})/\sqrt{2}$ are not the Wick-rotated version 
of the usual ones $(\Gamma_0 \pm \Gamma_{\psi})/\sqrt{2}$. 

As for the global symmetry of the action, however, 
it is important 
to recognize that if we redefine the spinor coordinates 
as $\theta^1 \rightarrow \theta^1, \theta^2 
\rightarrow \Pi\theta^2$, we can eliminate 
the matrix $\Pi$ from the fermion action (\ref{fermionaction}), 
since the kinetic term is invariant under this 
redefinition.  
Clearly, the continuous bosonic symmetry 
is now extended to 
SO(8) by treating $(x^i, y^i)$ as an SO(8) vector 
and also by assuming that the {\it redefined} spinor coordinates 
transform according to the usual transformation law, 
($\theta^I \rightarrow \exp (\omega^{ab}\Gamma_{ab}/4)\theta^I\equiv R\theta^I, \theta^I \Gamma_0 \rightarrow \theta^I\Gamma_0
R^{-1}) $
where $a, b, \ldots$ run over the directions transversal to 
$z, \psi$. Equivalently, if we insist on using the original 
spinor coordinates without making the redefinition, 
the SO(8) symmetry is hidden in such a way that 
$\theta^1$ and 
$\theta^2$   transform differently in general as given 
by 
\EQ
\theta^1 \rightarrow R\theta^1, \quad \theta^2 
\rightarrow \Pi R \Pi \theta^2. 
\EN
The manifest SO(4)$\times$SO(4) corresponds to 
$\Pi R \Pi =R$ for $R\in $SO(4)$\times$SO(4). 
   Naively, we expect that the RR field strength 
$F_5 = \epsilon_5 + \ast \epsilon_5$ breaks 
the symmetry down to SO(4)$\times$SO(4). However, 
the theory is actually SO(8) symmetric 
in the PP wave limit. The situation is that the presence of 
RR field is crucial for the emergence of 
the mass term for fermions,  but that PP-wave limit is 
not fully sensitive with respect to 
its directions. The spectrum 
of fluctuations consists of an 
infinite number of SO(8) multiplets 
when the states are suitably relabeled.  
In the following, unless stated otherwise, we will use the manifestly SO(8) symmetric 
conventions, by renaming the 4+4 bosonic vector 
$(x, y)$ as $x^i$ $(i=1, \ldots, 8)$ and by 
using the redefined fermion spinors $(\theta^1, \theta^2)$ 
after eliminating $\Pi$.  Once the world-sheet theory 
has SO(8) symmetry, it is reasonable to expect 
that the whole theory after second quantization 
should have the same symmetry.

\subsection{Quantization}
Let us next briefly treat the quantization of our 
Euclidean string theory. 
The appropriate mode expansion for bosons 
 is, suppressing the 
SO(8) indices,  
\EQ
x(\tau, \sigma) = {1\over \sqrt{2\alpha}}
\sum_{n=-\infty}^{\infty}{1 \over \sqrt{E_n}}
\Big(\e^{-E_n\tau +in\sigma/\alpha}a_n
+ \e^{E_n\tau-in\sigma/\alpha}a^{\dagger}_n\Big) , 
\EN
\EQ
p(\tau, \sigma) = {i \over 2\pi}{dx \over d\tau} = {i\over 2\pi\sqrt{2\alpha}}
\sum_{n=-\infty}^{\infty}\sqrt{E_n}
\Big(-\e^{-E_n\tau +in\sigma/\alpha}a_n
+ \e^{E_n\tau-in\sigma/\alpha}a^{\dagger}_n\Big), 
\EN
with 
 \EQ
E_n =\sqrt{1+{n^2 \over \alpha^2}} =\sqrt{1 + {R^4 n^2 \over J^2}}. 
\EN
The canonical commutation relations are  
\EQ
[x(\tau, \sigma) , p(\tau, \sigma')] =i 
\delta(\sigma-\sigma'), \quad \quad [a_n, a_m^{\dagger} ]=\delta_{nm} .
\EN
We emphasize that for nonzero $\tau$ the field variables 
$x(\tau, \sigma), p_x(\tau, \sigma)$ are not 
self-adjoint:
\EQ
x(\tau, \sigma)^{\dagger}=x(-\tau, \sigma). 
\EN
However, we used the adjoint notation $\dagger$ 
with understanding that after inverse Wick rotation 
these variables reduce to the standard ones.

For fermions, the mode expansions are 
\EQ
\theta^1 (\tau, \sigma)=
{1\over \sqrt{2\alpha}}\sum_{n=-\infty}^{\infty}
{1\over \sqrt{2E_n(E_n-{n \over \alpha})}}
\Big(\e^{-E_n \tau + in\sigma/\alpha}
b_n
+
\e^{E_n \tau - in\sigma/\alpha}
{i\over \sqrt{2}}\Gamma_+\Gamma_0 b_n^{\dagger}
\Big) ,
\EN
\EQ
\theta^2 (\tau, \sigma)=
{i\over \sqrt{2\alpha}}\sum_{n=-\infty}^{\infty}
{1
\over \sqrt{2E_n(E_n+{n \over \alpha}) }}
\Big(\e^{-E_n \tau + in\sigma/\alpha}b_n
-
\e^{E_n \tau - in\sigma/\alpha}
{i\over \sqrt{2}}\Gamma_+\Gamma_0b_n^{\dagger}
\label{theta2mode}
\Big) .
\EN
Because of massive nature, two spinors $\theta^1, \theta^2$ 
are not independent to each other 
after the imposition of the equations 
of motion. This is of course the most crucial 
difference between massive and massless fermions 
on the world sheet. 
 
The fermionic canonical commutation relations are  
\EQ
\{
\theta(\tau, \sigma), 
\Pi_{\theta}(\tau, \sigma')\}
=P_+\delta(\sigma-\sigma'), 
\EN
\EQ
\{\theta(\tau, \sigma), \theta(\tau, \sigma' )\}
=0=\{\Pi_{\theta}(\tau, \sigma), \Pi_{\theta}(\tau, \sigma')\}
,
\EN
\EQ
\{b_n, b_m^{\dagger}\} =P_+\delta_{nm} 
,
\EN
where the new fermion variables are 
defined by
\EQ
\theta(\tau, \sigma)\equiv \theta^1(\tau, \sigma) + 
i \theta^2(\tau, \sigma), \quad 
\overline{\theta}(\tau, \sigma)
\equiv\theta^1(\tau, \sigma) -i \theta^2
(\tau, \sigma)
,
\EN 
with  
$
\Pi_{\theta}(\tau, \sigma)\equiv 
{i \Gamma_0\Gamma_-\over 2\pi}\overline{\theta}.
$
Here we suppressed spinor indices and the 
Weyl projection for notational 
brevity. All the spinor indices in the present paper 
should be understood as with the Weyl projection,  
$\Gamma_{\mu} = h_- \Gamma_{\mu} h_+
$ or  
$\Gamma_{\mu} = h_+ \Gamma_{\mu} h_-
$ ($h_{\mp}=(1\mp\Gamma_{11})/2$), 
depending on the positions of $\Gamma$ matrices. 
Remember that in our convention, 
\EQ
(\Gamma_0\Gamma_{\pm})^T =\Gamma_0\Gamma_{\pm} , 
\quad \Gamma_0^T =-\Gamma_0, \quad \Gamma_{\pm}^T 
=\Gamma_{\pm},  \quad (\Gamma_0\Gamma_{\pm}\Gamma_{\mu\nu})^T
=-\Gamma_0\Gamma_{\pm}\Gamma_{\mu\nu} ,
\EN
and also that the spinor coordinates are further 
projected owing 
to the $\kappa$-gauge condition as 
\EQ
P_+b_n=b_n, \quad P_-b_n^{\dagger}=b_n^{\dagger} ,
\EN
with the  projection operators 
\EQ
P_{\pm}\equiv {1\over 2}\Gamma_{\pm}\Gamma_{\mp}, 
\quad \Gamma_{\pm}P_{\pm}=0=P_{\pm}\Gamma_{\mp} ,
\EN
satisfying 
\EQ
P_{\pm}^T =P_{\mp}, \quad 
P_+\Gamma_+ =\Gamma_+ , 
\quad 
\Gamma_- P_+
= \Gamma_- , \quad etc. 
\EN 
Below, the $\kappa$-gauge projection will also be mostly 
suppressed. 
Note that $P_{\pm}$ are hermitian and positive definite:
\EQ
P_{\pm}^{\dagger}=P_{\pm}={1\over 2}\Gamma_{\pm}\Gamma_{\pm}^{\dagger}.
\EN
The conjugation property of the fermionic coordinates 
is 
\EQ
\theta^I(\tau, \sigma)^{\dagger}=-{i\over \sqrt{2}}\Gamma_-\Gamma_0\theta^I(-\tau, \sigma). 
\EN

The total Hamiltonian 
is 
\[
H={1\over 2}\int_0^{2\pi \alpha} d\sigma\, 
:\Big(
2\pi p^2 + {1\over 2\pi} (x')^2 +{1\over 2\pi}x^2 
+{1\over 2\pi}
 \theta \Gamma_0\Gamma_-\theta'
+{1\over 2\pi}
 \overline{\theta} \Gamma_0\Gamma_-
\overline{\theta}'
+{i\over \pi} \theta \Gamma_0\Gamma_-
\overline{\theta}
\Big):
\]
\EQ
=
\sum_n E_n \Big(a_n^\dagger a_n 
+b_n^{\dagger}b_n
\Big) , 
\label{hamiltonian}
\EN
giving the Euclidean equations of motion
($
x(\tau, \sigma)=\e^{H\tau} x(0, \sigma) \e^{-H\tau}, etc 
$). 
We can define Fock space as usual on the 
basis of the Fock vacuum $|0\rangle$ and its conjugate, 
satisfying $a_n|0\rangle =b_n|0\rangle =0
=\langle 0|a_n^{\dagger}=\langle 0|b_n^{\dagger}$. 
The Hamiltonian  is then self-adjoint in the usual sense and 
positive definite with eigenvalues 
\EQ
E=\sum_n E_n(\sum_{i=1}^8 N_{i, n}^{bose} +
 \sum_{\alpha=1}^8 N_{\alpha, n}^{fermi}) ,
\EN
where $N_{i, n}^{bose}$ and $N_{\alpha, n}^{fermi}$ are 
the numbers of excitations for  eight bosonic 
directions and for eight independent 
fermionic directions, respectively. 
The theory is a perfectly well denfined 
Euclidean system with physical positivity 
(reflection positivity). 
The hidden SO(8) symmetry is manifest in 
the above final form of the 
Hamiltonian. 
To avoid confusion, it should be noted that the SO(8) 
symmetry cannot be interpreted as real 
rotations in the background space-time. 
For instance, the dilaton and axion are not singlet 
with respect to SO(8), although they are 
with respect SO(4)$\times$SO(4).  

The states where only the zero modes ($n=0$) are 
excited must correspond to the massless supergravity 
fields around the AdS background. 
Then, it is possible to interpret the 
eigenvalue of the Hamiltonian as 
the difference of field dimensions and 
angular momentum, $E =\Delta -J$, by 
identifying the ground states as the chiral primary 
states of the supergravity multiplets with the given $J$.  
This conforms to the semi-classical 
result  in the previous section. 
For each $J$, the states with only fermion zero modes are 
excited gives the basic 128 (bosons) + 128 (fermions) 
physical supergravity states.  The infinite tower 
on these states formed by exciting 
bosonic zero modes correspond to 
orbital fluctuations around the 
$\psi$ direction. Following ref. \cite{bmn}, it is 
natural to interpret 
the excitation energies $\sqrt{1+ {R^4n^2\over J^2}}  -1$ of nonzero modes as the anomalous dimensions of gauge-field 
operators corresponding to 
general string fields, in the approximation 
where string loop effects are ignored. 

It is 
more convenient to use $SO(8)$ spinor notations 
for Dirac matrices than 
the above 10 dimensional ones.  We briefly 
indicate the expressions  using them. 
The Hamiltonian and the canonical 
commutation relations are 
\EQ
H={1\over 2}\int_0^{2\pi\alpha}
d\sigma 
:\Big[
2\pi p^2 +{1\over 2\pi}(x')^2 +{1\over 2\pi}x^2
-{i\over 2\pi}(\theta\theta' +
\overline{\theta}\overline{\theta}') +{1\over \pi}\theta
\overline{\theta}
\Big]: ,
\EN
\EQ
\{\theta_a(\sigma), \overline{\theta}_b(\sigma')\}=2\pi
\delta_{ab}\delta(\sigma-\sigma') , \, \quad 
\theta(\tau, \sigma)^{\dagger} =
\overline{\theta}(-\tau, \sigma).
\EN
If we wish to return to the convention 
with the spinor factor $\Pi$ before 
our redefinition $\theta_2 \rightarrow \Pi\theta_2$, 
the natural canonical spinor coordinates which 
we denote by $\psi,  \overline{\psi}$ are 
related to our $\theta, \overline{\theta}$ 
by a canonical transformation breaking 
SO(8) to SO(4)$\times$SO(4), 
\EQ
\psi={1\over 2}(1+\Pi)\theta + {1\over 2}(1-\Pi)\overline{\theta}, \quad 
\overline{\psi}={1\over 2}(1-\Pi)\theta + 
{1\over 2}(1+\Pi)\overline{\theta}
\EN
where,  in the standard $(8\times8)$ 
SO(8) spinor notation for gamma matrices 
($(\gamma_i\gamma_j^T+
\gamma_j\gamma_i^T)_{ab}=2\delta_{ab}, 
\, \, (\gamma_i^T\gamma_j+
\gamma_j^T\gamma_i)_{\dot{a}\dot{b}}=2
\delta_{\dot{a}\dot{b}}$) 
\EQ
\Pi_{ab}=\Pi_{ba}=(\gamma_1
\gamma_2^T\gamma_3\gamma_4^T)_{ab}, \quad 
\Pi_{\dot{a}\dot{b}}=\Pi_{\dot{b}\dot{a}}=(\gamma_1^T
\gamma_2\gamma_3^T\gamma_4)_{\dot{a}\dot{b}}. 
\EN
In terms of these coordinates,  
 the Hamiltonian takes the 
form \EQ
H={1\over 2}\int_0^{2\pi\alpha}
d\sigma 
:\Big[
2\pi p^2 +{1\over 2\pi}(x')^2 +{1\over 2\pi}x^2
-{i\over 2\pi}(\psi\psi' +
\overline{\psi}\overline{\psi}') +{1\over \pi}\psi\Pi
\overline{\psi}
\Big]:
\EN

The standard dynamical   supersymmetry generators\cite{met,mettsey}, 
which only respect \\SO(4)$\times$SO(4), are 
\EQ
Q_{\dot{a}}^-= \int_0^{2\pi\alpha} d\sigma 
\Big(
(p\cdot\gamma-{i\over 2\pi}x\cdot\gamma\Pi)\psi
-{1\over 2\pi}x'\cdot \gamma \overline{\psi}
\Big)_{\dot{a}} ,
\label{qminus}
\EN 
\EQ
\overline{Q}_{\dot{a}}^-= \int_0^{2\pi\alpha} d\sigma 
\Big(
(p\cdot\gamma+{i\over 2\pi}x\cdot\gamma\Pi)  \overline{\psi}
-{1\over 2\pi}x'\cdot \gamma \psi
\Big)_{\dot{a}} , 
\label{qminusbar}
\EN 
in terms of the spinor coordinates $\psi, \overline{\psi}$. 
Nontrivial part of the supersymmetry algebra is 
\EQ
\{Q_{\dot{a}}^-, \overline{Q}_{\dot{b}}^-\}= 2H\delta_{\dot{a}\dot{b}}
+\sum_{(i, j)\in(1,2,3,4)}i
(\gamma_{ij}\Pi)_{\dot{a}\dot{b}}J_{ij}
-\sum_{(i, j)\in(5,6,7,8)}i
(\gamma_{ij}\Pi)_{\dot{a}\dot{b}}J_{ij} ,
\label{dynamicalalge}
\EN
\EQ
J_{ij}=\int_0^{2\pi\alpha}d\sigma
\Big(
x_i p_j -x_jp_i -{1\over 4\pi}i\psi 
\gamma_{ij}\overline{\psi}
\Big) .
\EN
More convenient representation of this algebra 
for our later purpose will be, suppressing spinor indices, 
\EQ
\{Q_1^-, Q_1^-\} 
=\{Q_2^-, Q_2^-\} = 2H
\EN
\EQ
\{Q^-_1, Q^-_2\}
= -\sum_{(i, j)\in(1,2,3,4)}
\gamma_{ij}\Pi J_{ij}
+\sum_{(i, j)\in(5,6,7,8)}
\gamma_{ij}\Pi J_{ij}
\label{susyalg2}
\EN
with 
\EQ
Q_{1}^-\equiv {1\over \sqrt{2}}( Q^- + \overline{Q}^-), 
\quad Q_2^-\equiv {1\over \sqrt{2}i}(Q^--
\overline{Q}^-) .
\EN

The susy algebra does not respect the SO(8) symmetry 
of the world-sheet action and hence of the Hamiltonian. 
This suggests that supersymmetry may not be  
perworful enough in constraining the dynamics of 
the system. It is tempting to introduce 
 fermionic symmetry generators respecting SO(8) as 
\EQ
R_{\dot{a}}^-= \int_0^{2\pi\alpha} d\sigma
\Big[(p-{i\over 2\pi}x)\cdot \gamma_{\dot{a}b}
\theta_{b}-{1\over 2\pi}x'\cdot \gamma_{\dot{a}b}
\overline{\theta}_b
\Big], \quad 
\overline{R_{\dot{a}}^-}=\Big(R_{\dot{a}}^-\Big)^{\dagger} ,
\EN
 using the manifestly SO(8) symmetric 
spinor coordinates $\theta, \overline{\theta}$. Indeed we can check that these `pseudo' susy generators commute 
with the Hamiltonian, and their anticommutators are 
\EQ
\{R_{\dot{a}}^-, \overline{R_{\dot{b}}^-}\}
=2\delta_{\dot{a}\dot{b}}H
-i\gamma^{ij}_{\dot{a}\dot{b}}L^{ij}
, \quad \{R_{\dot{a}}^-, R_{\dot{b}}^-\}=0 ,
\EN
with 
\EQ
L^{ij}=\int_{0}^{2\pi\alpha}d\sigma
\Big[
x^ip^j-x^jp^i +
{i\over 4\pi}\theta\gamma^{ij}
\overline{\theta}
\Big] . 
\label{L}
\EN
Because of the  plus sign of the fermionic 
contribution  in (\ref{L}), this algebra 
does not close with a finite number of generators.
\footnote{
In the original version of the present paper, 
an erroneous statement has given 
with respect to this point. }
Alternatively, if we combine the 
standard susy generators with our SO(8), 
the algebra is extended to an infinite 
dimensional algebra. This suggests that we can have 
much stronger constaints on the dynamics of the 
system by combining  
 SO(8) and susy than taking into account only the 
standard supersymmetry.  
In any case, it seems very important to further 
clarify the role of the hidden SO(8) symmetry. 

\vspace{0.4cm}
\section{Holography: Correspondence between string 
S-matrix and OPE}
\setcounter{equation}{0}
Once the free string theory is given, it should in principle be  
straightforward to construct superstring field theory 
following the procedure   in the flat case.  
In the case of PP-wave limit on the real null geodesic, 
this task  has been undertaken in \cite{spvolo}. 
We will pursue this subject in our Euclidean 
setting elsewhere. 
In the present work, instead of proceeding to such a 
direct construction of string field theory, we 
study the problem how the fundamental relation (\ref{relcorr}) 
of holography can be realized on the basis 
of the Euclidean string field theory from a more 
general standpoint,  
without assuming explicit 
expression of the string-field Hamiltonian.  
Instead, we assume that the set of physical string states 
correspond, at the boundary, to a complete (in 
the sense of OPE) set of gauge-theory 
operators with definite conformal dimensions,  
as proposed in \cite{bmn}. 
We are now initiating to study the general structure of `holographic' 
string field theory. 

\subsection{Euclidean S-matrix}
Let  the 
string-field theory Hamiltonian be ${\cal H}$,  
\EQ
{\cal H} = {\cal H}^{(0)} + {\cal H}^{(1)} + {\cal H}^{(2)} \cdots
\equiv {\cal H}^{(0)} + {\cal V}
,
\EN
where ${\cal H}^{(0)}$ is the free Hamiltonian 
corresponding to the single-string Hamiltonian (\ref{hamiltonian}) and 
${\cal H}^{(i)}$ $(i=1, 2, \ldots)$ are  
interaction vertices, expanded in powers 
of string coupling. If necessary we have to 
include some counter 
terms corresponding to renormalization effects when we 
study the string-loop effects. 
In terms of this Hamiltonian, the tunneling amplitudes  
after taking string interactions into account are 
essentially  described by matrix elements 
of the transition operator 
\[
U(\tau_2, \tau_1)=\exp \Big[-{\cal H}(\tau_2
-\tau_1)\Big], \quad (\tau_2>\tau_1)
\]
in the limit $\tau_2=T\rightarrow \infty, 
\tau_1=-T\rightarrow -\infty$. 
More precisely, the 
S-matrix is defined by multiplying the 
asymptotic transition operator for amputation of 
external lines, 
\EQ
S=\lim_{T\rightarrow \infty}
\e^{{\cal H}^{(0)}T}U(T, -T)\e^{{\cal H}^{(0)}T}.
\EN
Euclidean Schr\"{o}dinger equation corresponding to 
these transition operators is 
\EQ
{d \over d\tau}|\Psi_{in} (\tau)\rangle= -{\cal H}| \Psi_{in}(\tau)\rangle.
\EN
Here we put the 
subscript `{\it in}', since states obeying 
this equation are supposed to reduce to 
incoming asymptotic states in the 
limit $\tau\rightarrow -\infty$. 
For `{\it out}' states corresponding to 
$\tau \rightarrow +\infty$, we have similarly 
\EQ
{d \over d\tau}|\Psi_{out} (\tau)\rangle= {\cal H}| \Psi_{out}(\tau)\rangle.
\EN
These asymptotic states should corresponds to 
the various composite operators as identified by 
the work \cite{bmn}. 
Note that the sign here is chosen such that 
$\partial_{\tau} \sim \pm \Delta$ in the limit 
$\tau\rightarrow \pm \infty$ (See (\ref{clamp})). 
The choice of nonnormalizable boundary condition 
as discussed in section 3 corresponds to 
the behavior that $|\Psi_{in}(-T)\rangle $ 
and $|\Psi_{out}(T)\rangle $ exponentially 
increase in the limit $T \rightarrow \infty $ for generic 
states other than the ground state.

To incorporate the asymptotic Hamiltonian ${\cal H}^{(0)}$,  
we use the interaction representation:
\EQ
|\Psi_{in}^I(\tau)\rangle =
\e^{{\cal H}^{(0)} \tau} |\Psi_{in}(\tau)\rangle,
\EN
satisfying 
\EQ
|\Psi_{in}^I(\tau)\rangle =
U_+(\tau)|\Psi_{in}^I(-T)\rangle,
\EN
with 
\EQ
U_+(\tau)={\cal T}\exp\Big(-\int_{-T}^{\tau} d\tau {\cal V}_+(\tau)
\Big) , \quad 
{\cal V}_+(\tau)=\e^{{\cal H}^{(0)}\tau}{\cal V}
\e^{-{\cal H}^{(0)}\tau}. 
\EN
Similarly, 
we define 
\EQ
|\Psi_{out}^I(\tau)\rangle =
\e^{-{\cal H}^{(0)} \tau} |\Psi_{out}(\tau)\rangle
,
\EN
satisfying 
\EQ
|\Psi_{out}^I(\tau)\rangle =
U_-(\tau)|\Psi_{out}^I(+T)\rangle
,
\EN
with 
\EQ
U_-(\tau)=\overline{{\cal T}}\exp\Big(-\int^{+T}_{\tau} d\tau {\cal V}_-(\tau)
\Big) , \quad 
{\cal V}_-(\tau)=\e^{-{\cal H}^{(0)}\tau}{\cal V}
\e^{{\cal H}^{(0)}\tau}. 
\EN
Here ${\cal T}$ $(\overline{{\cal T}})$ are 
(anti) time-ordering operators with respect to $\tau$. 
The formal definition of S-matrix is then
\EQ
S=U_-(\tau)^{\dagger}U_+(\tau)\equiv 1+T, 
\EN 
which satisfies
\EQA
\langle \Psi^I_{out}(\tau)| \Psi_{in}^I(\tau)\rangle 
&=&\langle \Psi^I_{out}(+T)|
S|
\Psi_{in}^I(-T)\rangle \nonumber\\
&=&\langle \Psi^I_{out}(+T)|
 {\cal T} \exp\Big(-\int_{-T}^{+T}
d\tau {\cal V}_+(\tau)\Big)
|\Psi_{in}^I(-T)\rangle . 
\EQN 
Keep in mind that this definition amounts 
to normalizing two-point functions on the 
gauge-theory side as identity matrix.  
We here used the symbol $T$ for the nontrivial 
part of the S-matrix, 
since we do not expect any confusion owing to this 
abuse of notations. 
The perturbation expansion of the $T$ matrix 
is 
\[\hspace{-5.8cm}
\langle b| T|a \rangle
=\lim_{T\rightarrow \infty}
\Big[-\langle b|{\cal V}|a\rangle
\int_{-T}^{T}d\tau 
\e^{(E_b-E_a)\tau}
\]
\[
+\sum_c \langle b | {\cal V}|c\rangle \langle c|{\cal V}
|a \rangle 
\int_{-T}^{T}d\tau 
\e^{(E_b-E_c)\tau}
\int_{-T}^{\tau} d\tau_1\e^{(E_c-E_a)\tau_1}
\]
\[
-\sum_c\sum_d 
\langle b | {\cal V}|c\rangle \langle c|{\cal V}
|d \rangle \langle d|{\cal V}
|a \rangle 
\int_{-T}^{T}d\tau 
\e^{(E_b-E_c)\tau}
\int_{-T}^{\tau} d\tau_1\e^{(E_c-E_d)\tau_1}\]
\EQ\hspace{5cm}\times 
\int_{-T}^{\tau_1} d\tau_2\e^{(E_d-E_a)\tau_2}
 + \cdots
\Big] .
\label{tmatrix}
\EN
For notational simplicity  we assumed that the 
states are eigenstates of energy. In general, we have to 
superpose them. Remember also that the states 
$\langle b|$ and $|a\rangle$ here are 
those in the interaction representation. 

Now, we have to make  a remark which will be crucial 
 in our arguments below.  
In order to extract proper physical results from 
various  
 infinite time integrals involved 
in this expression,  
 it is necessary to 
carefully define the asymptotic {\it in} and {\it out} states 
 using wave packet formalism. Remember, for instance,  
that the physical justification of the familiar $i\epsilon$ prescription for ordinary Minkowskian S-matrix 
rests on the use of wave packets. 
In our case, because of confining harmonic potential 
in the transverse directions, it is not possible to 
make positions of wave packets 
far apart in the transverse directions. However, we can 
define asymptotic states along the $\psi$ direction. 
Wave packet formalism is perfectly applicable to the 
present system by regarding it effectively 
as a (1+1)-dimensional system. Namely, 
instead of considering states with a definite 
angular momentum ($J$) and energy, we have 
to superpose 
states with various 
different angular momenta around $J$ with 
appropriate weights. Then recalling 
that the exponent of the wave function 
with definite angular momentum is 
$E(J)\tau -J\psi$, where $\psi$ denotes 
of course the zero mode part 
which is not constrained by 
the Virasoro conditions (\ref{vira1}) and 
(\ref{vira2}), the velocity 
of a wave packet with average angular momentum $J$ moving along the large circle corresponding to the angle $\psi$ is 
determined by the saddle-point equation,\footnote{
The reader might wonder how to justify the 
wave packet after Wick rotation. We understand this 
in the formal sense of saddle-point approximation 
in the large $T$ limit: 
The classical trajectory is defined by Wick rotation. 
However, the superposition of states around the 
classical trajectory can be formed by integrating 
along appropriate directions in the complex $J$-
plane, such that 
the saddle-point approximation is justified.  
Similar situation arises when we consider 
tunneling in quantum mechanics. Although 
there is no wave packet of the ordinary sense 
in the tunneling region, 
we can still talk about wave packets for 
initial and final states of tunneling processes. 
}  
\EQ
v=R{dE\over dJ} . 
\EN
Note that this velocity is the correction to the 
velocity, with respect to the affine time $\tau$, of classical solution which is $R$ (=light 
velocity)  
in the same convention as we use here. 
Thus the relative velocities of wave packets are 
generically of order O$((\alpha^3R)^{-1})$, except for 
strictly supergravity states for which $v=0$. 
This means that two wave packets  with a generic initial distance of order O$(R)$ 
(standard `macroscopic' scale of the present background) requires 
a time interval of order O$(T)\sim $O$(R^2\alpha^3)$ 
to collide.   Conversely, if wave packets made a 
collision at $\tau\sim 0$, they become free 
 when average distances are 
of standard macroscopic scale after time passed as $\tau\sim$ O$(R^2\alpha^3)$. 

Now from known behavior of string scattering, 
we can assume that strings can interact `quasi-locally', 
in the sense that the matrix elements of interaction vertices 
are nonvanishing only for the wave packets 
which come close up to string scale ($\sim 1$ 
by our choice of length unit) when they collide.  
Although in the extreme high-energy limits 
strings exhibit a characteristic non-local behavior 
signified by a space-time uncertainty relation \cite{yoneuncertain},  
this assumption can be justified for generic 
finite-energy processes.  
We suppose that the initial ($|a\rangle$) and final 
($\langle b|$) states 
{\it in the interaction representation} consist of 
wave packets such that the distances among 
the packets are in general of macroscopic order $R$.  
 In order that the 
expression (\ref{tmatrix}) has nonvanishing 
contribution, the wave packets, superposed 
with the energy 
factors $\e^{\pm ET}$, must be 
such that the colliding wave packets 
come close to the order of string scale.  
There is a sign ambiguity here depending on 
how these states are prepared. 
We shall show that 
the choice of positive exponent $\e^{ET}$ 
is appropriate to realize  
the basic 
holographic relation. 

It is interesting to see 
what is the situation in the Schr\"{o}dinger 
picture. Owing to the relations 
\[
\e^{{\cal H}^{(0)}T}|\Psi_{in}^I (-T)\rangle 
=|\Psi_{in}(-T)\rangle, 
\quad 
\e^{{\cal H}^{(0)}T}|\Psi_{out}^I (T)\rangle 
= |\Psi_{out}(T)\rangle, 
\]
our choice amounts to the assumption that the distances among wave packets in the Schr\"{o}dinger picture  
are of order one 
at the boundary $\tau \sim \pm T$.  
In other words, in the Schr\"{o}dinger picture, 
the interaction occurs mostly near the 
boundaries. 
Note that once a definite sign is chosen, 
amplitudes with 
the opposite sign would vanish since they cannot 
represent wave packets coming close to each 
other at the right time 
(in the interaction picture) and hence 
the matrix elements of string vertices would be zero. 
Remember that the difference of 
positive and negative exponents corresponds to 
the time duration of order $2T$. Therefore, 
if states are prepared in such a way that 
the wave packets come close 
for one sign, the states with opposite sign 
would represents wave packets whose relative positions  are far apart 
from each other and hence the matrix elements 
would vanish. 

The crucial implication of this 
assumption is
that any initial $|a\rangle$ or final $\langle b|$ states 
{\it involving more than one string} 
can contribute to connected scattering amplitudes 
only when 
the energy exponents of all strings involved are 
positive.  
As we see later, this special feature, which is somewhat  peculiar from the viewpoint of ordinary S-matrix, 
is related to the fact that our initial and final 
states should correspond to 
the multiple products of composite operators 
at the boundaries $(\tau =
\pm T, \, T\rightarrow \infty)$  of the {\it single} tunneling  
null geodesic. Therefore the initial and final 
states necessarily represent certain short-distance 
limits of correlation functions into two groups. 

Armed with this general 
consideration, we can 
now discuss our main problem, holographic interpretation 
of the $T$-matrix elements (\ref{tmatrix}). 
In the present paper, we 
restrict ourselves within the tree approximation. 

\subsection{3-point amplitude}
The 3-point amplitudes come only from the 
first term in the expansion (\ref{tmatrix}), 
\EQ
 { \e^{(E_b-E_a)T}- \e^{-(E_b-E_a)T}\over
E_a-E_b }\, \langle b| {\cal H}^{(1)}|a\rangle .
\EN
Because of the conservation of $\alpha$ (or $J$), 
only one of the states $\langle b|$  or $|a\rangle$ 
can be a two-particle state. According to the 
rule established in the previous subsection, the part 
contributing to the S-matrix in the large time limit is 
either 
\EQ
{\e^{(E_b-E_a)T}\over E_a-E_b}\, \langle b| {\cal H}^{(1)}|a\rangle
\quad \mbox{or} \quad 
{\e^{(E_a-E_b)T}\over E_b-E_a}\, 
\langle b| {\cal H}^{(1)}|a\rangle, 
\EN
depending on whether $b$ $(1\rightarrow 2)$ or $a$ 
($2\rightarrow 1$)  
is the two-particle state, 
respectively.\footnote{
There is a subtlety here. The energy denominator 
$1/(E_a-E_b)$ becomes singular at $E_a=E_b$. 
However, we consider generic processes of 
wave packet states 
with $E_a\ne E_b$.  If necessary we can take the 
limit $E_a\rightarrow E_b$ after the limit 
of large $T$. This subtlety is related with 
the similar problem for `extremal' correlators 
 in the ordinary AdS-Gravity/CFT correspondence 
without the PP-wave limit. 
See, {\it e.g.},  \cite{slee}. 
}
Thus, in terms of conformal dimensions, 
 3-point functions (namely, matrix elements 
of ${\cal H}^{(1)}, 
\langle b| {\cal H}^{(1)}|a\rangle 
=V_{ijk}
$) in general take the form 
\EQ
{V_{ijk}\over (\Delta_k-\Delta_i -\Delta_j)}
\e^{(\Delta_i +\Delta_j-\Delta_k)T},
\EN
using light-cone 3-point vertex $V_{ijk}$ 
where the pair of indices $(i, j)$ designates 
the two-particle state,  either $|a\rangle $ (initial state) 
or $\langle b|$ (final state),  and 
$k$ being the single particle state. 
By relating the large 
time interval $T$ with the short-distance cutoff length by 
\EQ
\e^{-T}\sim |x_i-x_j| 
\EN
in conformity  with the semiclassical result (\ref{clamp}), 
this precisely corresponds to the short distance limit 
$|x_i-x_j|\rightarrow 0 \, \, (T\rightarrow \infty) $ of 
the 3-point function  $\langle O_i(x_i)O_j(x_j)O_k(x_k)\rangle$, 
provided the two-point functions are normalized to 1, 
which  essentially amounts to setting $|x_i-x_k|=|x_j-x_k|=1$ 
in this expression.  
Due to the assumption of the 
completeness of BMN operators with definite conformal dimensions corrresponding to 
the transverse oscillations of string,  the indices 
$(i, j, k, \ldots)$ of OPE coefficients behave essentially as the R-symmetry 
indices of SO(8),  and the short distance OPE, which 
is a main assumption for the boundary theory, 
takes the form
\EQ
O_i(0)O_j(x)\sim \sum_k {1\over |x|^{\Delta_i+\Delta_j-\Delta_k}}C_{ijk}O_k(0), \quad 
x\rightarrow 0.
\EN 

Thus the 3-point vertex is expressed in terms of the 
OPE coefficient as 
\EQ
V_{ijk}=(\Delta_k-\Delta_i-\Delta_j)C_{ijk}. 
\EN
This form has been conjectured in \cite{constetal}, but now is 
directly explained as a consequence of holography. 
We note that  some computations 
supporting this form of 3-point vertex have been 
reported in the 
case of the ordinary real-time formulation. 
For a (partial) list of such works, see \cite{3point}
(
See, however, the note added at the end of the present paper. 
). 
The explicit construction of string field 
theory vertex using the formalism 
developed in the previous section 
is left as a future work. 

We can rewrite this as a commutation relation,  
\EQ
{\cal H}^{(1)} =\pm[{\cal H}^{(0)}, {\cal C}]. 
\label{h1comm}
\EN 
We use this operator  
notation in the tree approximation in the sense that,  
in computing matrix elements of this relation, 
we only allow intermediate states which in terms of 
Feynman diagrams represents tree diagrams.  
The operator ${\cal C}$ is defined such that its matrix 
elements give the OPE coefficient $C_{ijk}$. 
No confusion should arise here since 
${\cal H}^{(0)}$ does not change the particle number. 
The sign in (\ref{h1comm}) is 
$+ (-)$ when $a$ ($b$) is the two-particle state, corresponding 
to $2\rightarrow 1$  ($1\rightarrow 2$) process. 
In the more standard notation, we would write ${\cal C}$ as 
${1\over 2}(\Psi^{\dagger})^2C\Psi$ or ${1\over 2}\Psi^{\dagger}C(\Psi)^2$, 
corresponding to 
$1\rightarrow 2 $ or $2\rightarrow 1$, respectively. 

As has already been noticed in ref. \cite{verlinde} 
in a different context, 
this form is consistent with the susy algebra. 
By second-quantizing the algebra (\ref{susyalg2}), 
we must have in the present approximation 
\EQ
2{\cal H}^{(1)}=\{{\cal Q}_1^{-(0)}, {\cal Q}_1^{-(1)}\}+
\{{\cal Q}_1^{-(1)}, {\cal Q}_1^{-(0)}\}=
\{{\cal Q}_2^{-(0)}, {\cal Q}_2^{-(1)}\}
+\{{\cal Q}_2^{-(1)}, {\cal Q}_2^{-(0)}\}, 
\EN
and 
\EQ
\{{\cal Q}_1^{-(0)}, {\cal Q}_2^{-(1)}\}+
\{{\cal Q}_2^{-(0)}, {\cal Q}_1^{-(1)}\}=0.
\label{q1q2order1}
\EN
Combining the lowest order 
algebra with (\ref{h1comm}), these relation determine the first order susy operators as 
\EQ
{\cal Q}_1^{-(1)}=\pm [{\cal Q}_1^{-(0)}, {\cal C}], 
\quad 
{\cal Q}_2^{-(1)}=\pm [{\cal Q}_2^{-(0)}, {\cal C}] 
\EN
with the requirement of SO(4)$\times$SO(4)  invariance, 
\EQ
[{\cal J}_{ij}, {\cal C}]=0 ,
\label{so8symm}
\EN
which comes from (\ref{q1q2order1}), as it should be 
since SO(4)$\times$SO(4) is a kinematical symmetry of the system. Of course, it is possible to 
require higher global symmetry SO(8). 

\subsection{4-point amplitude}
We now have to study both the first and second terms 
in (\ref{tmatrix}). 
Let us first consider the contribution from the second term
which is after the integration
\[\hspace{-9.8cm}
\sum_c\langle b|{\cal H}^{(1)}
|c\rangle
\langle c|{\cal H}^{(1)}|a\rangle 
\]
\EQ\hspace{2cm}
\times \Big[
{\e^{(E_b-E_a)T}\over (E_a-E_b)(E_a-E_c)}
+{\e^{(E_a-E_b)T}\over (E_b-E_a)(E_b-E_c)}
+{\e^{(E_a+E_b-2E_c)T}\over 
(E_c-E_b)(E_c-E_a)} 
\Big] .
\label{4point}
\EN
Note that in our tree approximation 
the intermediate states $|c\rangle $ contain only one 
particle which is not connected to initial or final 
states directly. In other words, if $|c\rangle$ 
is a multi-particle 
state, such particles are `spectators' except for a single 
particle corresponding to the internal line of 4-point Feynman diagram. 
There are two cases which we have to consider 
separately.
\begin{enumerate}
\item  Both $a\sim (i, j)$ and $b\sim (k, \ell)$ 
are two-particle states.
Our rule then tells us that the third term in (\ref{4point}) 
should correspond to a 4-point correlator 
\[
\langle O_i(x_i)O_j(x_j)O_k(x_k)O_{\ell}(x_{\ell})\rangle
,
\]
in the short-distance limit 
$|x_i-x_j|\sim |x_k-x_{\ell}|
\rightarrow \e^{-T}$. 
This is indeed realized by the relation established 
in the previous subsection 
which shows that the third term is given as 
\EQ
\sum_m C_{ijm}C_{mk\ell}\, \, \e^{(\Delta_i+\Delta_j 
+\Delta_k +\Delta_{\ell}-2\Delta_m)T}
\Leftarrow \e^{(E_a+E_b)T}\sum_c\e^{-2E_cT}\langle b|{\cal C}|c\rangle 
\langle c|{\cal C}|a\rangle, 
\EN
where $m$ corresponds to the internal line. 
Remembering  that two-point functions are 
normalized to be 1 in our convention, 
this is the correct OPE result for the 
4-point correlator, assuming that the 
transverse string states correspond to 
the complete set of gauge-theory operators with 
definite conformal dimensions. 

We note that 
the case where the external lines in initial and final 
states cross does not contribute according to 
our rule, since in that case we would have 
negative exponents for some of the external lines 
owing to the energy factor $\e^{-2E_cT}$.   
It is a general rule that the intermediate state 
with energy factor $\e^{-2ET}$ must always be 
associated with 
a single internal line without any spectator 
strings. Even if we include such terms by relaxing the 
restriction in preparing the initial and final states, 
they would be nonleading contributions in the limit 
$T\rightarrow \infty$. The amplitudes we are considering 
thus correspond generically to that in Fig. 1.

\item  Either $a$ or $b$ is a three-particle state $(i, j, k)$. 
In this case, the second or first term, respectively,
 should contribute to  the S-matrix. However, neither 
does take the precise structure expected from the 
short distance OPE of three operators, except for the 
energy exponent which is correct. 
For definiteness, let us 
consider the case where $b$ is the 3-particle state 
({\it i.e.} $1\rightarrow 3$ process).  
Using the same notational convention 
introduced in the case of 3-point amplitude, 
the first term takes the form 
\EQ
-{\e^{(E_b-E_a)T}\over E_a-E_b}\, \, \langle b|
[{\cal H}^{(0)}, {\cal C}]{\cal C}|a\rangle.
\EN
The energy exponent takes the required form 
$E_b-E_a=\Delta_i+\Delta_j +\Delta_k -\Delta_{\ell}$. 

The solution of this problem is that there must be the 
contribution from the 4-point vertex ${\cal H}^{(2)}$ 
in this case. Indeed if we assume 
\EQ
{\cal H}^{(2)} \Rightarrow -{\cal C}[{\cal H}^{(0)}, {\cal C}]
\EN
for $1\rightarrow 3$ (from right to left) 4-point vertex, 
the total contribution to the present 4-point amplitude is 
\EQ
-{\e^{(E_b-E_a)T}\over E_a-E_b}\, \,\langle b|
({\cal C}[{\cal H}^{(0)}, {\cal C}]+[{\cal H}^{(0)}, {\cal C}]
{\cal C}
)
|a\rangle
=\e^{(E_b-E_a)T}\, \langle b| {\cal C}{\cal C}|a \rangle ,
\EN
as is required for the correct short distance OPE 
for the present case.  To avoid confusion, 
we note that in the matrix-element form,  the above 
expressions involving commutators are 
\EQ
\langle b|{\cal C}
[{\cal H}^{(0)}, {\cal C}]|a\rangle=\sum_{P(ijk)}\sum_m(E_m+E_k-E_{\ell})
C_{ijm}C_{mk\ell} ,
\EN
\EQ
\langle b|
[{\cal H}^{(0)}, {\cal C}]{\cal C}|a\rangle=\sum_{P(ijk)}\sum_m(E_i+E_j-E_m)
C_{ijm}C_{mk\ell}, 
\EN
where the summation symbol $\sum_{P(ijk)}$ 
should be understood as summation over 
inequivalent pertition of $(i, j, k)$ into 
1+2 groups.  
Note that because of the conservation of angular momentum we can always replace the energy 
$E$ by conformal dimension $\Delta$ in the 
commutator.

If we repeat the same argument for the case 
where $a$ is the 3-particle state, the second term of 
(\ref{4point}) is now 
\EQ
{\e^{(E_a-E_b)T}\over E_b-E_a}\, \, \langle b|
{\cal C}[{\cal H}^{(0)}, {\cal C}]|a\rangle. 
\label{3to1}
\EN
Hence the $3\rightarrow 1$ 4-point vertex must be 
$
{\cal H}^{(2)} \Rightarrow [{\cal H}^{(0)}, {\cal C}]{\cal C},  
$
since it contributes to the 4-point amplitude as 
\[
{\e^{(E_a-E_b)T}\over E_b-E_a}\,  \langle b|
[{\cal H}^{(0)}, {\cal C}]{\cal C}|a\rangle, 
\]
and gives the correct 4-point amplitude 
$\e^{(E_a-E_b)T}
\langle b|{\cal C}{\cal C}|a\rangle $, combining with (\ref{3to1}). 
 
\end{enumerate}

\noindent
Our analysis shows that to reproduce the OPE structure corresponding to 
two-group short distance limit of conformal operators it is necessary to 
add the second order interaction terms ${\cal H}^{(2)}$ corresponding to $1\rightarrow 3$ and 
$3\rightarrow 1$ matrix elements. The one corresponding to 
$2\rightarrow 2$ is not necessary. Or more appropriately, 
we should say that the requirement of holography cannot directly 
fix the form of $2\rightarrow 2$ matrix element 
within the present tree approximation.  
This is essentially owing to 
our special choice in preparing the initial and final wave packet states, 
since according to our prescription the negative energy exponents 
associated with the multi-particle states can be ignored.

Our next task is  to 
study whether the 4-point interaction vertex obtained 
above from the requirement of holographic 
correspondence is 
consistent with the susy algebra. 
The algebra to be satisfied is 
\[
2{\cal H}^{(2)}=
\{{\cal Q}_1^{-(1)}, 
 {\cal Q}_1^{-(1)}\}+\{{\cal Q}_1^{-(0)}, 
{\cal Q}_1^{-(2)}\}
+\{{\cal Q}_1^{-(2)}, 
{\cal Q}_1^{-(0)}\}
\]
\EQ=
\{{\cal Q}_2^{-(1)}, {\cal Q}_2^{-(1)}\}+\{{\cal Q}_2^{-(0)}, 
{\cal Q}_2^{-(2)}\} 
+\{{\cal Q}_2^{-(2)}, 
{\cal Q}_2^{-(0)}\}, 
\EN
\EQ
\{{\cal Q}_1^{-(1)}, {\cal Q}_2^{-(1)}\} 
+\{{\cal Q}_1^{-(0)}, 
{\cal Q}_2^{-(2)}\} 
+\{{\cal Q}_2^{-(0)}, 
{\cal Q}_1^{-(2)}\} + \mbox{transposed expressions}=0 .
\EN
It should be kept in mind that in dealing with these algebras 
in the present section 
we are considering only the matrix elements of 
the types,  either $(1\rightarrow n)$ or $(n\rightarrow 1)$, 
separately. In the present tree approximation, 
this restriction of matrix elements can be implemented 
consistently. 
Other types of 
matrix elements cannot be treated since we have 
not determined the vertices of general types. 
It is easy to check that all of these relations are 
satisfied for ${\cal H}^{(2)} = -{\cal C}[{\cal H}^{(0)}, {\cal C}]$ 
$ (1\rightarrow 3)$ 
or for $[{\cal H}^{(0)}, {\cal C}]{\cal C}$ $(3\rightarrow 1)$ 
if we choose the susy generator of the present order 
as 
\EQ
{\cal Q}^{-(2)}_{1, 2}
=-{\cal C}[{\cal Q}^{-(0)}_{1,2}, {\cal C}] 
\quad 
\mbox{or}\quad 
{\cal Q}^{-(2)}_{1, 2}
=[{\cal Q}^{-(0)}_{1,2}, {\cal C}]{\cal C} ,
\EN
respectively, providing that the SO(8) symmetry relation (\ref{so8symm}) is valid. 
The holography and  supersymmetry conforms to 
each other  in a quite miraculous way.  

\begin{center}
\begin{figure}
\begin{picture}(140,120)
\put(120,20){\epsfxsize 220pt 
\epsfbox{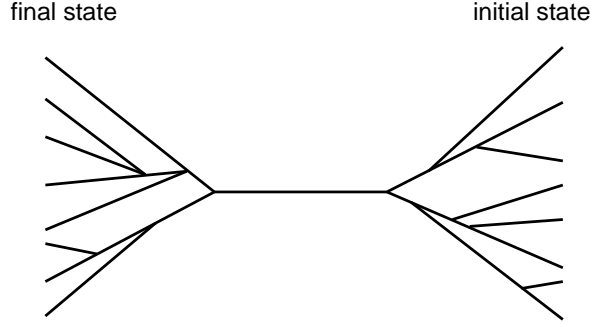}}
\end{picture}
\caption{ This shows a generic diagram we are considering 
with our choice of wave packet states. } 
\label{Fig1} 
\end{figure}
\end{center}

\subsection{5-point amplitude} 
The integration in the third term of (\ref{tmatrix}) gives 
\[
\hspace{-5cm}
\sum_{c, d}\langle b|{\cal H}^{(1)}|c\rangle 
\langle c|{\cal H}^{(1)} |d\rangle 
\langle d|{\cal H}^{(1)}|a\rangle
\]
\[
\times \Big[{\e^{(E_b-E_a)T} \over (E_a-E_b)
(E_a-E_d)(E_a-E_c)}
+{\e^{-(E_b-E_a)T} \over (E_b-E_a)
(E_b-E_d)(E_b-E_c)}
\]
\EQ
+{\e^{(E_b+E_a-2E_c)T}\over (E_c-E_b)(E_c-E_d)(E_c-E_a)}
+{\e^{(E_b+E_a-2E_d)T}\over (E_d-E_b)(E_d-E_a)(E_d-E_c)}
\Big]
.
\EN

\vspace{0.3cm}
\noindent
Comparing with the previous cases, we see that 
among these four terms the first or second term 
represent $(1\rightarrow 4)$ or $(4\rightarrow 1)$ 
process, respectively, and the 3rd or 4th term 
represent $(3\rightarrow 2)$ or $(2\rightarrow 3)$ 
process, respectively. 
Note in particular that in all other cases 
there arise negative exponents for multi-string 
initial or final states. 
Also, we have to take into account 
the contribution of higher-point vertices coming from 
second and first terms in (\ref{tmatrix}).

We first consider the case $(1\rightarrow 4)$. 
By adding the contributions of higher-point vertices, 
total amplitude is 

\[\hspace{-6cm}
{\e^{(E_b-E_a)T}\over E_a-E_b}
\Big[
\sum_{c,d}
{\langle b|{\cal H}^{(1)}|c\rangle 
\langle c|{\cal H}^{(1)} |d\rangle 
\langle d|{\cal H}^{(1)}|a\rangle
\over 
(E_a-E_d)(E_a-E_c)}
\]
\EQ
+\sum_d {\langle b|{\cal H}^{(2)}|d\rangle 
\langle d|{\cal H}^{(1)}|a\rangle
\over 
E_a-E_d}
+\sum_c {\langle b|{\cal H}^{(1)}|c\rangle 
\langle c|{\cal H}^{(2)}|a\rangle
\over 
E_a-E_c}
+\langle b|{\cal H}^{(3)}|a\rangle
\Big]
.
\EN
Substituting the previous results for ${\cal H}^{(1)}$ and  
${\cal H}^{(2)}$, this expression is reduced to 
\[
{\e^{(E_b-E_a)T}\over E_a-E_b}\langle b|
\Big(
-[{\cal H}^{(0)},{\cal C}]{\cal C}{\cal C}-{\cal C}
[{\cal H}^{(0)}, {\cal C}]{\cal C}
+{\cal H}^{(3)}\Big)|a\rangle .
\]
Thus we see that holography requires us to set  
\EQ
{\cal H}^{(3)}
\Rightarrow -{\cal C}{\cal C}[{\cal H}^{(0)}, {\cal C}], 
\label{3rdh13}
\EN
by which the $1\rightarrow 4$ amplitude 
takes the correct form 
$
\e^{(E_b-E_a)T}\langle b|{\cal C}{\cal C}{\cal C}|a\rangle. 
$

The $4\rightarrow 1$ case is just the transpose 
(hermitian conjugate) of this 
process so that we find
\EQ
{\cal H}^{(3)}\Rightarrow [{\cal H}^{(0)}, {\cal C}]
{\cal C}{\cal C}. 
\label{3rdh31}
\EN

As is always the case in the present paper, 
the multiplication symbol 
must be understood  within the tree 
approximation. Note that if there are 
spectator strings they just pass through 
the matrix elements of ${\cal C}$ as identity. 
Thus these products represent 
`monotonic'  flow of either increasing or 
decreasing number of strings. 

Let us next turn to the case $2\rightarrow 3$. In this case, 
the total amplitude is by taking into account the 
second order vertex ${\cal H}^{(2)}$,  
\EQ
\e^{(E_b+E_a)T}\, 
\Big[
\sum_{c, d}\e^{-2E_dT}{\langle b|{\cal H}^{(1)}|c\rangle 
\langle c|{\cal H}^{(1)} |d\rangle 
\langle d|{\cal H}^{(1)}|a\rangle
\over 
(E_d-E_b)(E_d-E_a)(E_d-E_c)}
+\sum_d\e^{-2E_dT}{\langle b|{\cal H}^{(2)}|d\rangle 
\langle d|{\cal H}^{(1)}|a\rangle 
\over (E_d-E_b)(E_d-E_a)} 
\Big]  .
\EN
On using the previous results, this is reduced to 
\[
-\sum_d
\e^{(E_b+E_a-2E_d)T}\, {\langle b|[{\cal H}^{(0)} ,{\cal C}{\cal C}]|d\rangle 
\over (E_d-E_b)}\langle d|{\cal C}|a\rangle.
\]
This gives the correct expression 
$
\sum_d \e^{(E_b+E_a-2E_d)T}\langle b|{\cal C}{\cal C}|d\rangle
\langle d|{\cal C}|a\rangle 
$
for the $2\rightarrow 3$ correlation function in terms of the 
OPE coefficients. 
The transposed case $3\rightarrow 2$ can be treated 
in the same way. As in the case of 4-point case, the interaction 
vertices corresponding to   
 $2\rightarrow 3$ and $3\rightarrow 2$ do not contribute to the 
leading short-distance behavior 
according to our prescription. 

Finally, by examining the susy algebra to the 3rd order 
using the above results, we find that the 3rd order 
term, 
\EQ
{\cal Q}_{1,2}^{-(3)}\Rightarrow -
{\cal C}{\cal C}[{\cal Q}_{1,2}^{-(0)},{\cal  C}], 
\EN
for $1\rightarrow 4$ and their hermitian conjugate 
$[{\cal Q}_{1,2}^{-(0)},{\cal  C}] {\cal C}{\cal C}$ for 
$4\rightarrow 1$ satisfy all the required properties 
in the similar way as we have found for lower orders. 

\subsection{General case}
We can continue this analysis to arbitrarily higher 
orders.   In fact, we can easily guess the general 
forms from the above results. To reproduce the 
leading short-distance behaviors of the 
correlators,  higher-interaction vertices (or contact terms) 
are necessary for the 
type $1\rightarrow n+1$ and their transpose $n+1\rightarrow 1$ 
$(n\ge 1)$ and are given as 
\EQ
{\cal H}^{(n)}_{(1\rightarrow n+1)}
=-{\cal C}^{n-1}[{\cal H}^{(0)}, {\cal C}], 
\quad 
{\cal H}^{(n)}_{(n+1\rightarrow 1)}
=[{\cal H}^{(0)}, {\cal C}]{\cal C}^{n-1}. 
\EN
We can prove this by a simple induction. It is 
sufficient to consider the case $1\rightarrow n$. 
       The general form of perturbative expansion 
of the T-matrix is 
\[
{\e^{(E_b-E_a)T}\over E_a-E_b}
\sum_{c_1, c_2, \ldots, c_{n-2}}
{\langle b|{\cal V}|c_{n-2}\rangle 
\langle c_{n-2}|{\cal V}|c_{n-3}\rangle \langle c_{n-3}|
{\cal V}|c_{n-4}\rangle 
\cdots \langle c_1|{\cal V}|a\rangle \over 
(E_a-E_{c_{n-2}})(E_a-E_{c_{n-3}})\cdots 
(E_a-E_{c_1})}\]
\[
\equiv {\e^{(E_b-E_a)T}\over E_a-E_b}
\langle b|{\cal T}_{n-1}|a \rangle ,
\]
where ${\cal V}$ can be any vertices up to the order 
${\cal H}^{(n-1)}$. Suppose that the sum in this expression is shown to be 
of the form 
\[
{\cal T}_{n-1}=-[{\cal H}^{(0)}, {\cal C}^{n-1}]
\]
for the cases $n\le N$ with the above form of   the 
vertices. 
This means that 
\EQAN
\langle b|{\cal T}_{N}|a \rangle &=&-\sum_{k=1}^{N}
\sum_{c_{N-1}}{\langle b|{\cal H}^{(k)}|c_{N-1}\rangle
\langle c_{N-1}|[{\cal H}^{(0)}, {\cal C}^{N-k}]|a\rangle
\over E_a-E_{c_{N-1}}}\\
&=&\sum_{k=1}^{N}
\sum_{c_{N-1}}\langle b|{\cal H}^{(k)}|c_{N-1}\rangle
\langle c_{N-1}|{\cal C}^{N-k}|a\rangle .
\EQNN
Holography requires that this expression 
is equal to 
\[
-\langle b| [{\cal H}^{(0)}, {\cal C}^N]|a\rangle. 
\]
which requires 
${\cal H}^{(N)}=-{\cal C}^{N-1}[{\cal H}^{(0)}, {\cal C}]$, 
completing the induction. 

We now have a Hamiltonian 
for our Euclidean and `holographic' 
string field theory, once the 3-point 
OPE coefficients ${\cal C}$ are fixed, 
which is exact at least at the 
tree level approximation in computing holographic 
S-matrix elements of the type in Figure 1: 
\EQ
{\cal H}={\cal H}^{(0)} -
{1\over 1-{\cal C}}[{\cal H}^{(0)}, {\cal C}]
+[{\cal H}^{(0)}, {\cal C}]{1\over 1-{\cal C}} ,
\EN
where the first interaction terms should be understood 
for the matrix elements $1\rightarrow n+1$ and 
the second for $n+1\rightarrow 1$. 
 
The susy generators corresponding to the above form 
of the general interaction vertices are 
\EQ
{\cal Q}^{-(n)}_{1,2 (1\rightarrow n+1)}
=-{\cal C}^{n-1}[{\cal Q}_{1,2}^{-(0)}, {\cal C}], 
\quad 
{\cal Q}^{-(n)}_{1,2 (n+1\rightarrow 1)}
=[{\cal Q}_{1,2}^{-(0)}, {\cal C}]{\cal C}^{n-1}.
\EN
These expressions can be rewritten in such a way that 
the susy algebra is manifestly satisfied:
\EQ
{\cal H}={1\over 1-{\cal C}}{\cal H}^{(0)}(1-{\cal C}), 
\quad 
{\cal Q}_{1,2}^{-}={1\over 1-{\cal C}}{\cal Q}_{1,2}^{-(0)}
(1-{\cal C}),
\EN
for $1\rightarrow n+1$ and  their transpose  
for $n+1\rightarrow 1$, respectively, 
for now $n\ge 0$. 
Remarkably, the full interacting operators of 
holographic string-field theory  
are obtained from the 
free theory by a similarity transformation composed 
of the OPE coefficients at least in the present 
classical (tree level) context. 
Of course 
the kinematical generators, which 
commute ${\cal C}$  
are not affected by this similarity transformations. 

We emphasize that even if the interacting 
Hamiltonian is obtained by a similarity 
transformation, the system is not 
trivial. S-matrix is nontrivial in our approach by 
its fundamental assumption. There is no contradiction 
here. Remember that, already at the level of 3-point amplitude, the form of the vertex $[{\cal H}^{(0)}, {\cal C}]$,  
which is of course the infinitesimal form of the 
above similarity transformation,  
would indicate vanishing of on-shell 3-point 
amplitude in ordinary Minkowski theory, 
because of energy conservation. 
This was not the case in our Euclidean theory, 
since we have the denominator of energy difference 
$1/(E_a-E_b)$, instead of delta function $\delta(E_a-E_b)$. 
What we have established for higher vertices are the 
extension of this phenomenon to 
higher orders.

\subsection{Exact form of the Hamiltonian?}
As has been already warned, for the matrix elements 
of the more general type as $m\rightarrow n$ $(m, n\ne 1)$,  
our prescription cannot fix their forms, since 
their matrix elements 
of wave packet states vanish when they appear in the 
amplitudes with our prescription. 
So there are two options as for the existence of 
other types of interaction vertices. 
\begin{enumerate}
\item There are no other types than those 
corresponding to $1\rightarrow n$ or $n\rightarrow 1$. 
\item There are other types as well. 
Holography alone is 
not sufficient to fix them.  
\end{enumerate}
The question is then what is the criterion in 
deciding the correct direction. A possible  
criterion which we can think of seems to require 
supersymmetry as a quantum symmetry beyond 
the tree approximation. 
The first option seems  unlikely to be true, 
since it is difficult to satisfy the susy algebra 
with only those types of interaction vertices. 
An important task is now to try to 
construct the other terms assuming the 
possibility 2 above. This requires us to 
quantize the string field theory and hence 
is beyond the scope of the present paper. 
We hope to discuss this question elsewhere.

\vspace{0.4cm}
\section{Concluding remarks}
In the present paper, we have investigated 
the question how 
holographic principle for 
the string theory in AdS$_5\times$S$^5$ 
background should be formulated in the context 
of PP-wave limit. 
We started from some puzzles involved in the 
important proposal made in \cite{bmn}. 
We argued that all the puzzles mentioned are 
naturally resolved if we recall that the 
holographic correspondence as signified 
in the basic relation (\ref{relcorr}) 
between string theory and super Yang-Mills theory 
is correctly interpreted as a tunneling phenomena 
from the viewpoint of semiclassical approximation. 
This led us to consider a tunneling null geodesic 
as the basis for quantizing string theory to describe 
interactions of strings with large angular momenta 
along a particular direction of S$^5$. 
By assuming consistency of the string interaction 
with holography on the basis of tunneling 
picture, we found that the form of 
string vertices are strongly constrained in a 
way conforming to supersymmetry. 
 
There are several points which are left untouched and/or 
remain to be further clarified. 
First, we have not discussed the 
effect of string loops. 
We hope that our formalism 
serves as a foundation for discussing 
higher genus effects and its connection with 
SYM. 
Second, we have not performed concrete
 comparison of our results on string S-matrix with 
higher-point correlators on the SYM side. 
In particular, in our 
formulation, the cutoff parameter in 
forming short distance OPE has 
been assumed always to be the single quantity $\e^{-T}$ 
with one and the same normalization. 
However, on gauge-theory side, it is not trivial 
to achieve this in taking short-distance limit 
of general multiple products of operators in usual ways. 
For $1\rightarrow n$ or $n\rightarrow 1$ up to, at least,  
$n=5$, this can be achieved in flat 4-dimensional 
base space by setting all the distances equal for arbitrary 
pairs of $n$ points where the 
operators of initial or final states are located. However, for $n>6$, such simple 
configurations are not possible, as long as we assume the  
flat metric. One possible way out of this problem 
might be to introduce nontrivial conformal metric 
for the base space of boundary theory and identify 
the cutoff parameter using the distances defined by 
metrics with some nontrivial Weyl factor, 
such that we can assume a single cutoff $\e^{-T}$. 
In the limit $T\rightarrow \infty$ and $n\rightarrow \infty$, such metric would 
become very chaotic, but final physical results 
may show regular behaviors in the sense of our results.  
In the literature, 
some ambiguity of perturbative 4-point functions in taking 
short-distance limit 
has been reported \cite{beisert}.  We also mention that 
there have been some discussions \cite{chu} 
on the closure of OPE of BMN operators and its relevance to 
higher-point amplitudes. In view of our results, 
it is very important to develop a systematic 
method for constructing 
two-group short distance limits of multi-trace operators 
on gauge-theory side.  

With respect to the possible ambiguity of normalization, 
we have to keep in mind the following subtlety.  For string field-theory 
S-matrix, we have to sum over all 
different tree diagrams, while in OPE 
tree diagrams which are connected by 
channel duality must be equivalent and are not 
summed over, {\it provided} we have a 
complete set of operators. 
In our string-theory language, 
this `channel duality' would be 
an automatic consequence.  If these statements are justified, 
we have to normalize our initial and final states in comparing them with 
the CFT correlators, in 
such a way that the  degeneracy caused by 
the sum over dual channels are canceled. 
That amounts to dividing by the factor 
$(2n-3)!!$ for each $n$-particle initial or final states, 
which is equal to  the number of $(n+1)$-point Feynman diagrams 
in $\phi^3$ field theory. 

With the understanding of all these possible 
subtleties with respect to regularization and normalization, 
our results are essentially summarized by the following 
symbolic `reduction formula' 
connecting string S-matrix $S^E(T)$ with sufficiently large Euclidean time 
duration $2T$ 
with two-group correlators of SYM, 
\EQ
\langle i_1, i_2, \ldots, i_p| S^E(T)|j_1, j_2, \ldots, j_q\rangle_{conn} 
=\langle (\overline{O}_1\overline{O}_2\cdots \overline{O}_p)_T(1)
(O_1O_2\cdots O_q)_T(0)\rangle_{conn}
,
\EN
where the first and second groups of operators in the right-hand side 
are suitable short-distance limits of multiple products 
of BMN operators with negative and positive angular 
momentum, respectively,  with cutoff length $\e^{-T}$. 
One might wonder how about more general correlators 
with multi-centers. In principle, it should be possible to 
generalize our results to multi-center cases by introducing 
coherent states with respect to the zero-mode 
operators to shift the positions of wave packets along 
transverse directions.  
However, we cannot settle this question at this stage 
unless we have at hand systematic 
methods of dealing with short-distance limit as discussed above. 
All these are left for the future works. 

To conclude, we add more remarks on some 
salient features in our arguments and results. 
\begin{enumerate}
\item Wave packet for correlation functions:

We had to assume wave packets in order to extract 
S-matrix elements correctly such that they 
precisely reproduce 
the behaviors of CFT correlators defined on the 
boundary. The reader may wonder what wave packets  
mean for the correlators. 
For this question, we should recall that on the 
 boundary we are treating nonstandard 
composite operators which involve 
infinitely large number of fields at the same space-time 
point in the form
$
\Tr[\cdots Z^J \cdots](0) 
$. 
In our way of establishing the holographic 
correspondence of BMN operators with the S-matrix of Euclidean 
string-field theory, we have to further take 
short-distance products of such operators. 
$
\Tr[\cdots Z^{J_1} \cdots](0) 
\Tr[\cdots Z^{J_2} \cdots](\epsilon) \cdots  
 \quad (\epsilon 
\sim \e^{-T}).
$
These are certainly very singular objects. In particular, 
by rewriting symbolically  as $Z=|Z|
\e^{i\psi}$, these are products of 
`plane waves' $\e^{iJ\psi}$ which have 
infinite extensions in the {\it configuration} space of 
fields. Remember that since we are taking 
the  limit $J\rightarrow \infty$ and treating 
$\alpha$ as a continuous parameter, 
the $\psi$ direction is now effectively noncompact. 
Then, it is a natural procedure to 
regularize the effect of the infinite extendedness 
in the configuration space of fields, by making 
wave packets even for correlators as we are forced to do 
in defining Euclidean S-matrix. 
It would be interesting to pursue this idea further. 
This might also be related to the resolution of possible 
ambiguity mentioned above. 

\item Connection with holographic renormalization group:

In our formulation, the identification of time parameter 
$\tau$ along a tunneling null geodesic 
with $\Delta -J$ was completely manifest. 
In particular, the 
cutoff parameter   $\e^{-T}$ 
is identified with  
the radial coordinate $z\sim e^{-T}$ at the boundary. 
In this sense, the Schr\"{o}dinger equation 
of string field theory can be interpreted as 
a version of holographic renormalization group 
equation. A very interesting problem now 
is whether it is possible to derive the same 
equation directly at the boundary using only the 
logic of super Yang-Mills theory. 
It is natural to expect that a Wilsonian approach 
to the renormalization group 
for the BMN operators would lead to such a 
derivation. 

\item Universality of holographic correspondence: 

An important question related to our approach 
is   `How universal is the 
holographic correspondence ?'.  For example, we 
may ask, as has already been alluded to before, 
whether we can extract the 
string degrees of freedom in Matrix theory 
in a similar fashion. 
Since we have formulated the holographic correspondence 
entirely using the Poincar\'{e} patch 
of the AdS geometry, there is  in principle no difficulty 
in  extending our method to such cases. For instance, it 
would be interesting to see how the results found in 
\cite{sekinoyo} for supergravity modes on the 
basis of picture proposed in \cite{jeyo} 
are generalized to stringy degrees of freedom.

\item Exact solvability of quantum string theory: 

We have presented strong evidence that 
holographic string theory has 
the  simple Hamiltonian which 
is obtained by a simple similarity transformation 
from the free theory at least in the 
tree approximation for a 
special class of interaction vertices. 
 We may try to construct a fully quantized 
unitary operator  for our 
string field theory at least formally 
as a generalization of our classical similarity 
transformation. 
This seems quite suggestive 
    from the viewpoint of seeking for nonperturbative 
definition of string/M theory.  

 \end{enumerate}

\vspace{0.3cm}
\noindent
{\it Note added}

\vspace{0.1cm}
\noindent
Responding to the comment from the referee, we would like to 
add a couple of remarks on the status of our prediction 
eq. (4.20) for further clarification. 
After the submission of the present paper, 
it has been pointed out in ref. \cite{pear}
that some of the previously reported 
results on the properties of the 
prefactor of 3-point vertex were groundless. In particular, 
the claim of some works in \cite{3point} 
that the prefactor of 3-point vertex of 
the (real time) string field theory is equal to the difference of 
energies is not valid, because they are based on a sign error,  
as pointed out in \cite{Pan}. 
However, this does not invalidate our 
prediction eq. (4.20) for several reasons:
\begin{enumerate}
\item Our claim is not that the 
prefactor itself is the difference of energies, but only 
that the correct 3-point function (OPE coefficient) on the 
gauge-theory side should take the form $C_{ijk}=V_{ijk}/
(\Delta_k-\Delta_i-\Delta_j)$ in terms of the string field 
vertex $V_{ijk}$. For this, it is not necessary 
that the prefactor itself takes the form of energy 
difference.  
\item In order to make comparison with computations 
on the gauge-theory side, we have to construct the 
Euclidean string field theory according to the formalism 
developed in section 3.  
In fact, from what we have discussed in section 4, 
the requirement of susy algebra in the approximation 
of classical string field theory is not sufficient to 
fix the interaction vertex uniquely, 
since we can always construct the 3-point interaction vertex formally 
such that it satisfies 
the first order form of the susy algebra, 
once a ${\cal C}$ preserving kinematical 
symmetries is given. The results would strongly 
depend on the choice of ${\cal C}$.  
For example, there is an ambiguity regarding to what extent possible 
kinematical symmetries 
(for instance, either
SO(4)$\times$SO(4)($\times$ Z$_2$) or 
SO(8)$\times$SO(8)) are assumed. In the 
standard treatment, even Z$_2$ symmetry is 
not taken into account as has been emphasized in \cite{chu2}. 
 Of course, 
for the result to be well behaved, 
the CFT coefficients should not diverge for the special 
cases when the energy is conserved. In other words, 
the 3-point vertex must have vanishing 
matrix elements for energy conserving 
processes. 
In the case of supergravity approximation, this 
is indeed satisfied before taking the 
plane-wave limit, as discussed in the literature 
in  the  name of `extremal' correlators \cite{slee} 
as we have mentioned in the text. 
It seems that all the known results for CFT 
coefficients from perturbative 
computations on the Yang-Mills side are 
also consistent with this property \cite{chuk}.  
\item On the gauge-theory side, it is known that 
we have to take into account operator mixing 
even at the lowest order level in order to 
maintain the standard form of the 
3-point function, which is the basic 
assumption, eq. (4.19),  in our arguments in section 4. 
In connection with this,  we would like to 
mention a new proposal, appeared after the submission 
of the present paper, in ref. \cite{chuk} 
for the prefactor such that it is indeed consistent with 
our prediction (4.20), 
compared with all the known perturbative 
results on gauge theory side, taking into account the operator mixing. 
More generally, however, 
we have to keep in mind an important 
question whether  or not 
the logic of  light-cone string field theory alone involves the first principle for constructing interacting string 
theory in curved space-time uniquely. 
Our consideration seems to indicate that the answer 
to this question is negative. 

\end{enumerate} 
We are currently working on the 
explicit construction of  string 
field theory which meets our criterion to be the 
`holographic string field theory'. We hope to report on 
this progress in a forthcoming paper. 

\vspace{0.5cm}
\noindent
{\large Acknowledgements}

The work by T. Y. is dedicated to the memory of 
Prof. Bunji Sakita. T. Y. has been continually 
inspired and encouraged by his passion and 
attitude to physics. The present work was completed  
while one (T. Y.) of the authors was visiting 
Brown University. He would like to thank Physics Department of Brown University for its hospitality and 
to Antal Jevicki and Horatiu Nastase for discussions. 

The present work is supported in part by Grant-in-Aid for Scientific 
Research (No. 12440060 and No. 13135205)  from the Ministry of  Education,
Science and Culture. 

\vspace{1cm} 
\noindent
{\large Appendix}
\renewcommand{\theequation}{A.\arabic{equation}}
\appendix 
\newcommand{\p}{\partial}
\newcommand{\pa}{\partial_a}
\newcommand{\xp}{x^+}
\newcommand{\xm}{x^-}
\newcommand{\txp}{\tilde x_+}
\newcommand{\txm}{\tilde x_-}
\newcommand{\tX}{\tilde X_i}
\newcommand{\tP}{\tilde \Phi}
\newcommand{\tp}{\tilde \psi_i}

\setcounter{equation}{0}

\vspace{0.3cm}
\noindent
In this appendix we show how the free massive world-sheet action
derived in section 3 can be obtained by first making
a suitable coordinate transformation and then expanding
the transformed action around the classical solution 
or the null-geodesic line. 

We first choose the coordinate transformation
so that one of the coordinates lies along the tunneling null-geodesic line 
of section 3, 
which makes it easy to expand the action around the solution 
since all coordinates of the solution 
are zero except for one along the geodesic.
The desired transformation is given by
\begin{eqnarray}
z &=&\frac{1}{\cosh \xp}\, , \\
r &=&\tanh \xp
-\xm+\frac{X^2}{2}\tanh \xp +\frac{\Phi^2}{2\tanh \xp}
+\frac{\Phi}{\sinh \xp}\, ,\\
x^i&=&\frac{X^i}{\cosh \xp}\qquad (i=1,2,3)\, ,\\
\psi&=&\xp+\frac{ \Phi}{\sinh \xp}\, ,\\
\theta&=&\Theta\, ,
\end{eqnarray}
following the method of ref. \cite{blauetal}.

After transforming the bosonic action by this map,
we expand it around a classical solution
$\xp=\tau$ and $\xm=X^i=\Phi=\Theta=0$. With
the coordinate expansion such as
\begin{eqnarray}
\xp=\tau+\frac{1}{R} \xp_{(1)}+\frac{1}{R^2} \xp_{(2)}+\cdots,\quad
\xm=\frac{1}{R} \xm_{(1)}+\frac{1}{R^2} \xm_{(2)}+\cdots,\quad
\textrm{etc},
\end{eqnarray}
we obtain the quadratic action 
\begin{eqnarray}\label{secondorderaction_in_app}
&&\hspace{-5pt}S={1\over 4\pi}\int d\tau \int_0^{2\pi \alpha} d\sigma 
\, 
\Big[
-2\p\xp\p\xm
+\Phi^2+\sum_{i=1}^{3}(X^{i})^2+\sum_{i=1}^{4}(Y^{i})^2 \\ \nonumber 
&&\hspace{-5pt}+
(\p\Phi)^2+\sum_{i=1}^{3}(\p X^i)^2+\sum_{i=1}^{4}(\p Y^i)^2
+2\frac{\cosh^3\tau}{\sinh^2 \tau}\Phi\dot\xm
-2\frac{\cosh^2\tau}{\sinh \tau}\p\Phi\p\xm
+{\cosh^2 \tau}(\p\xm)^2
\Big],
\end{eqnarray}
where we have 
omitted the subscript 
and introduced the coordinates $Y^{i}$
as the Cartesian coordinates defined by
$(\p \Theta)^2+\sin^2\Theta(\p \Omega_3)\sim
(\p \Theta)^2+\Theta^2(\p \Omega_3)= 
\sum_{i=1}^4(\p Y^i)^2$.

The extra term appearing in (\ref{secondorderaction_in_app}) 
can be dropped by virtue of the Virasoro constraints,
whose bosonic part is expanded to the quadratic order as
\begin{eqnarray}
&&-2R\dot\xm_{(1)}-2\dot\xm_{(2)}
-2\{\dot\xp_{(1)}\dot\xm_{(1)}-(\xp_{(1)})'(\xm_{(1)})'\}
+\Phi^2+\sum_{i=1}^{3}(X^{i})^2+\sum_{i=1}^{4}(Y^i)^2\\ \nonumber
&&+\dot\Phi_{(1)}^2-\Phi_{(1)}^{'2}
+\sum_{i=1}^{3}\{(\dot X^i_{(1)})^2-(X^i_{(1)})^{'2}\}
+\sum_{i=1}^{4}\{(\dot Y^i_{(1)})^2-(Y^i_{(1)})^{'2}\}\\ \nonumber
&&+2\frac{\cosh^3\tau}{\sinh^2 \tau}\Phi_{(1)}\dot\xm_{(1)}
-2\frac{\cosh^2\tau}{\sinh \tau}
(\dot\Phi_{(1)}\dot\xm_{(1)}-\Phi_{(1)}'x_{(1)}^{-'})
+\cosh^2 \tau(\dot x^{-2}_{(1)}-x^{-'2}_{(1)})+{\cal O}
\left(\frac1R\right)=0,
\end{eqnarray}
and
\begin{eqnarray}
&&-2Rx^{-'}_{(1)}-2x^{-'}_{(2)}
-\dot\xp_{(1)} x_{(1)}^{-'}
-x_{(1)}^{+'}\dot\xm_{(1)} 
+\dot\Phi_{(1)}\Phi_{(1)}'
+\sum_{i=1}^{3}\dot X_{(1)}^i X_{(1)}^{i'}
+\sum_{i=1}^{4}\dot Y_{(1)}^i Y_{(1)}^{i'}\\ \nonumber
&&\hspace{20pt}+\frac{\cosh^3\tau}{\sinh^2 \tau}\dot\Phi_{(1)} x^{-'}_{(1)}
-\frac{\cosh^2\tau}{\sinh \tau}
(\dot\Phi_{(1)}x^{-'}_{(1)}+\Phi_{(1)}'\dot x^{-}_{(1)})
+\cosh^2 \tau\dot\xm_{(1)}x^{-'}_{(1)}+{\cal O}\left(\frac1R\right)=0.
\end{eqnarray}
As is easily seen, solving the above constraints recursively 
in terms of $1/R$ gives $\dot \xm_{(1)}=x^{-'}_{(1)}=0$.
Substituting them into (\ref{secondorderaction_in_app}) 
and redefining $(\Phi,X^{i})$ as $X^i(i=1,\cdots,4)$, 
we finally obtain the the same free massive string action 
as (\ref{finalbosonicaction}):
\begin{eqnarray}
&&\hspace{-5pt}S={1\over 4\pi}\int d\tau \int_0^{2\pi \alpha} d\sigma 
\, 
\Big[
\sum_{i=1}^{4}(X^{i})^2+\sum_{i=1}^{4}(Y^{i})^2 
+\sum_{i=1}^{4}(\p X^i)^2+\sum_{i=1}^{4}(\p Y^{i})^2
\Big].\label{finalbosonicaction_in_app}
\end{eqnarray}
We can repeat this analysis to fermions and 
find the same result as given in section 3.

\end{document}